\documentclass[default]{aastex701}

\begin{document}

\title{DragonflyPol: Wide-Field Optical Linear Polarimetry with the Dragonfly Telephoto Array\\
(Instrument Description and Commissioning)}

\author[0000-0001-8749-1436]{Mehrnoosh Tahani}
\affiliation{Department of Physics \& Astronomy, University of South Carolina, Columbia, SC 29208, USA}
\email{mtahani@sc.edu}
\author[0000-0001-6480-5260]{Leo Hollberg}
\affiliation{Department of Physics and Department of Geophysics, Stanford University, Stanford, California 94305, USA}
\email{leoh@stanford.edu}
\author[0000-0001-6156-238X]{Hiroshi Akitaya}
\affiliation{Astronomy Research Center, Chiba Institute of Technology, 2-17-1 Tsudanuma, Narashino, Chiba 275-0016, Japan}
\affiliation{Hiroshima Astrophysical Science Center, 1-3-1 Kagamiyama, Hiroshima University, Higashihiroshima, Hiroshima 739-8526, Japan}
\email{akitaya@perc.it-chiba.ac.jp}
\author[orcid=0000-0002-0432-6847]{Jaeyeon Kim}
\affiliation{School of Space Research, Kyung Hee University, 1732 Deogyeong-daero, Yongin-si, Gyeonggi-do 17104, Korea}
\affiliation{Department of Astronomy \& Space Science, Kyung Hee University, 1732 Deogyeong-daero, Yongin-si, Gyeonggi-do 17104, Korea}
\email{jyeonkim@khu.ac.kr}
\author[0000-0002-2406-7344]{Deborah Lokhorst}
\affiliation{NRC Herzberg Astronomy \& Astrophysics Research Centre,
5071 West Saanich Road, 
Victoria, BC V9E2E7, Canada}
\email{deborah.lokhorst@nrc-cnrc.gc.ca}
\author[0000-0002-4542-921X]{Roberto Abraham}
\affiliation{Department of Astronomy \& Astrophysics, University of Toronto, 50 Saint George Street, Toronto, ON M5S 3H4, Canada}
\affiliation{Dragonfly Focused Research Organization, 150 Washington Avenue, Santa Fe, NM 87501, USA}
\email{roberto.abraham@utoronto.ca}
\author[0000-0002-8282-9888]{Pieter van Dokkum}
\affiliation{Department of Astronomy, Yale University, New Haven, CT 06520, USA}
\affiliation{Dragonfly Focused Research Organization, 150 Washington Avenue, Santa Fe, NM 87501, USA}
\email{pieter.vandokkum@yale.edu}
\author[0000-0003-4381-5245]{William P. Bowman}
\affiliation{Department of Astronomy, Yale University, New Haven, CT 06520, USA}
\affiliation{Dragonfly Focused Research Organization, 150 Washington Avenue, Santa Fe, NM 87501, USA}
\email{william.bowman@dragonfly1000.com}
\author[0009-0003-0123-3017]{Vishwa Koshene Gamage}
\affiliation{Department of Physics \& Astronomy, University of South Carolina, Columbia, SC 29208, USA}
\email{koshenev@email.sc.edu}
\author[0009-0009-4369-355X]{Paras Regmi}
\affiliation{Department of Physics \& Astronomy, University of South Carolina, Columbia, SC 29208, USA}
\email{pregmi@email.sc.edu}
\author[0000-0001-8746-6548]{Yasuo~Doi}
\affiliation{Department of Earth Science and Astronomy, Graduate School of Arts and Sciences, The University of Tokyo, 3-8-1 Komaba, Meguro, Tokyo 153-8902, Japan}
\email{doi@ea.c.u-tokyo.ac.jp}
\author[0000-0001-6099-9539]{Koji S. Kawabata}
\affiliation{Hiroshima Astrophysical Science Center, 1-3-1 Kagamiyama, Hiroshima University, Higashihiroshima, Hiroshima 739-8526, Japan}
\email{kawabtkj@hiroshima-u.ac.jp}

\collaboration{all}{(DragonflyPol and Dragonfly Teams)}

\correspondingauthor{Mehrnoosh Tahani}
\email{mtahani@sc.edu}

\begin{abstract}

We present DragonflyPol, a wide-field optical linear polarimetry capability implemented on the Dragonfly Telephoto Array. DragonflyPol leverages Dragonfly's modular, multi-lens architecture to obtain simultaneous measurements in four linear polarization orientations ($0^\circ$, $45^\circ$, $90^\circ$, and $135^\circ$) across a $\sim5\,deg^2$ field of view, distributed across 44 polarized lens--detector units. Four additional units serve as unpolarized reference channels. We describe a broad range of science goals enabled by this capability, including magnetic field mapping and tomography, dust grain properties, CMB foreground characterization, and the three-dimensional structure of diffuse interstellar clouds. We integrate Canon polarizers and Baader Sloan $r'$ bandpass filters into the drop-in filter holders of the Canon lenses, and conduct a three-phase laboratory characterization program to select optimal polarimetric components, measure contrast ratios and transmission efficiencies, and determine and mark the transmission axis of each polarizer with sub-degree repeatability. Laboratory measurements across all 44 deployed polarizers yield a mean noise-subtracted contrast ratio of $1228 \pm 104$ and a single-polarizer transmission efficiency of $\sim$33\% in the $r'$ band. On-sky commissioning, including twilight flat-field characterization and twilight-sky polarization measurements, confirms throughput stability across polarization groups and successful recovery of the expected Rayleigh scattering signal. DragonflyPol achieved first polarimetric light in 2025 September. 

\end{abstract}

%% Keywords should appear after the \end{abstract} command. 
%% PASP uses Unified Astronomy Thesaurus (UAT) concepts:
%% https://astrothesaurus.org
%% You will be asked to selected these concepts during the submission process
%% but this old "keyword" functionality is maintained in case authors want
%% to include these concepts in their preprints.
%%
%% You can use the \uat command to link your UAT concepts back its source.

% \keywords{\uat{Classical Novae}{251} --- \uat{Ultraviolet astronomy}{1736} --- \uat{History of astronomy}{1868} --- \uat{Interdisciplinary astronomy}{804}}

%% From the front matter, we move on to the body of the paper.
%% Sections are demarcated by \section and \subsection, respectively.
%% Observe the use of the LaTeX \label
%% command after the \subsection to give a symbolic KEY to the
%% subsection for cross-referencing in a \ref command.
%% You can use LaTeX's \ref and \label commands to keep track of
%% cross-references to sections, equations, tables, and figures.
%% That way, if you change the order of any elements, LaTeX will
%% automatically renumber them. 

\section{Introduction}

\begin{figure}[t]
    \centering
    \includegraphics[scale=0.423]{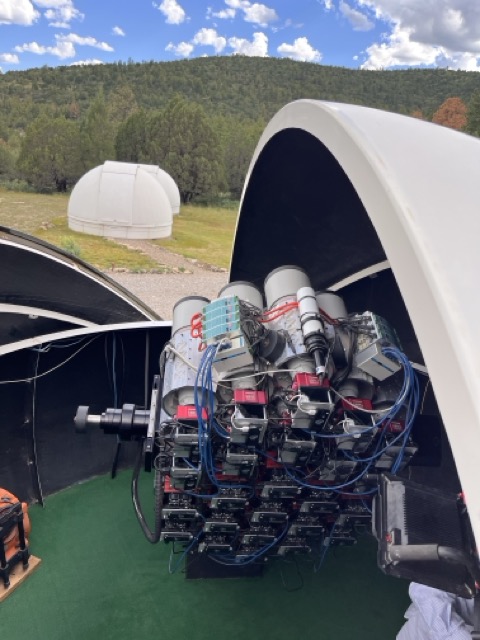}
     \caption{The Dragonfly Telephoto Array (Mount 1), showing the 24 Canon 400\,mm $f/2.8$ telephoto lenses with the DragonflyPol polarizer and bandpass filter assemblies installed. Image credit: DragonflyPol Collaboration.}
         \label{fig:Dragonfly}
\end{figure}

The Dragonfly Telephoto Array \citep{AbrahamvanDokkum2014} is an all-refractive mosaic telescope consisting of 48 commercial Canon 400\,mm $f/2.8$ telephoto lenses arranged in two clusters of 24 units, each hosted on a dedicated robotic tracking mount (Figure~\ref{fig:Dragonfly}). 
Its modular design enables rapid instrumentation upgrades by allowing each lens--detector unit to host distinct optical elements (e.g., filters or polarizers) while maintaining co-alignment and simultaneous imaging. The lenses incorporate nano-fabricated sub-wavelength anti-reflection coatings that suppress internal light scattering by an order of magnitude relative to conventional reflecting telescopes \citep{AbrahamvanDokkum2014}. The reduced scattering and modular design give Dragonfly exceptional sensitivity to low surface brightness structures over a $\sim$5\,deg$^2$ field of view. Since its inception, Dragonfly has produced a series of discoveries in the low surface brightness universe \citep[e.g.,][]{Cohenetal2018, vanDokkumetal2019, vanDokkumetal2022Natur, Liuetal2023}.

Dragonfly's architecture offers several properties that are directly advantageous for wide-field optical polarimetry. Its modular design allows individual lens--detector units to be equipped with different polarizer orientations, enabling simultaneous measurement of multiple linear polarization states across the full field of view in a single exposure. This capability is difficult to achieve with conventional single-aperture telescopes. The all-refractive optical design is particularly well-suited to polarimetry, as refractive optics with nano-fabricated anti-reflection coatings suppress scattering and minimize the systematics caused by internal reflections. In contrast, internal reflections in mirror-based telescopes introduce instrumental polarization that must be carefully characterized and corrected. Furthermore, Dragonfly's exceptional sensitivity to low surface brightness emission means that the polarized signal from faint diffuse structures (such as Galactic cirrus) can be detected in practical integration times. The use of commercial telephoto lenses also simplifies polarimetric integration, as the drop-in rear filter slot of each lens provides a natural, repeatable mounting point for polarizer and filter assemblies without modification to the optical path. Together, these properties make Dragonfly a natural platform for a wide-field polarimetric facility, motivating the development of DragonflyPol, described in this paper. 

Several optical polarimetry facilities and surveys have recently advanced the field, including RoboPol \citep{Blinovetal2023RoboPol}, DIPOL-2 \citep{Piirolaetal2014DIPOL2}, and the ongoing PASIPHAE survey \citep{Tassisetal2018}, which aims to map interstellar magnetic fields at high Galactic latitudes using a 1.3\,m telescope with a $30' \times 30'$ field of view. DragonflyPol and PASIPHAE complement each other: PASIPHAE targets high stellar density fields, while DragonflyPol's large field of view (roughly 20 times larger than PASIPHAE) and exceptional surface brightness sensitivity make it uniquely suited to diffuse, low surface brightness structures such as Galactic cirrus and extended molecular cloud envelopes where individual stars are sparse. These diffuse environments are precisely where some of the most pressing open questions lie: how magnetic fields are organized in the low-density interstellar medium, how dust grain alignment varies across different physical conditions, and how large-scale Galactic magnetic structure connects to the small-scale fields threading star-forming regions. No existing wide-field optical polarimetric facility combines Dragonfly's field of view, surface brightness depth, and simultaneous multi-angle polarization measurement capability.

DragonflyPol implements wide-field optical linear polarimetry on the Dragonfly Telephoto Array by equipping each lens--detector unit with a fixed linear polarizer and bandpass filter assembly. Polarizers are assigned orientations of 0$^\circ$, 45$^\circ$, 90$^\circ$, and 135$^\circ$ in repeating groups across the array, enabling simultaneous measurement of all linear Stokes parameters in a single exposure. The groups are assigned for calibration purposes.  The instrument achieved first light in September 2025, and commissioning observations are described in this paper. Calibration methodology, polarimetric performance characterization, and standard star results are presented in a companion paper (Tahani et al. in preparation). This paper is organized as follows. Section~\ref{sec:science} describes the science goals motivating DragonflyPol. Section~\ref{sec:overview} provides an overview of the Dragonfly Telephoto Array architecture. Section~\ref{sec:lab} presents the laboratory characterization program used to select polarizer components and determine their transmission axis orientations. Section~\ref{sec:instrumentCommission} describes the instrument configuration and on-sky commissioning observations. Section~\ref{sec:discussion} summarizes our conclusions.

\section{Science Goals}
\label{sec:science}
DragonflyPol's wide field of view, exceptional surface brightness sensitivity, and simultaneous multi-angle polarization capability enable a broad range of science goals, described below. Several of these goals, that have been difficult to access with existing facilities, are uniquely addressed by DragonflyPol's combination of field coverage and depth. 
\paragraph{Magnetic Field Mapping and Tomography}
Starlight polarization by magnetically aligned dust grains traces the plane-of-sky component of interstellar magnetic fields along the line of sight. When combined with stellar distances from \textit{Gaia}, these measurements enable tomographic reconstruction of the three-dimensional magnetic field geometry of the interstellar medium \citep[e.g.,][]{Panopoulouetal2019, Doietal2021, Pelgrimsetal2024}. To date, complete 3D magnetic field vectors in 3D space (6D maps) have been reconstructed in only two interstellar regions, incorporating both line-of-sight~\citep{Tahanietal2018} and plane-of-sky magnetic field observations and providing 6D magnetic maps of the Orion A and Perseus molecular clouds \citep{Tahanietal2022O, Tahanietal2022P}. These 6D studies demonstrate that magnetic fields preserve the imprint of large-scale Galactic fields and trace cloud formation and evolution, as previously suggested in line-of-sight field studies of \cite{HanZhang2007}. DragonflyPol's wide field of view and surface brightness sensitivity make it well suited to extend such analyses to additional regions, particularly diffuse environments and cloud envelopes where stellar densities are lower and existing narrow-field instruments are less effective. 

\paragraph{Cloud and Star Formation}
Magnetic fields play a significant role in the star-formation process. The orientation of magnetic fields with respect to density structures has important consequences for star formation \citep{Pattleetal2023PP7}. For example, magnetic fields parallel to an elongated filamentary cloud would resist the radial collapse of the cloud, while perpendicular fields would resist its azimuthal collapse.  Understanding the roles that magnetic fields play in this process necessitates mapping both the 3D clouds and their 3D magnetic field vectors~\citep{Tahani2022}. Additionally, understanding how magnetic fields guide the transition from diffuse to dense gas is central to theories of molecular cloud formation and stellar birth. Previous multi-wavelength polarimetric studies of the Orion A and Perseus molecular clouds have revealed how large-scale magnetic field geometry connects to small-scale star-forming structures \citep{Tahanietal2022O, Tahanietal2022P} and how these field morphologies reveal cloud formation history. These polarimetry studies enabled the identification of superbubbles and bubbles that had shaped the magnetic field geometry and influenced cloud formation. In the case of the Perseus molecular cloud, the reconstructed field lines revealed a structure invisible in total-intensity emission that had shaped the field and influenced the cloud's formation history; the presence of this structure was subsequently confirmed by independent kinematic observations \citep{Kounkeletal2022}.  However, extending these analyses to a broader sample of clouds and environments requires wide-field coverage and starlight polarimetry. DragonflyPol's large field of view enables simultaneous mapping of magnetic field morphologies across clouds, bubbles, shells, and their surrounding diffuse envelopes~\citep{Tahanietal2026}, connecting the fields threading star-forming regions to the larger Galactic environment and identifying their formation and evolution history.

\paragraph{Dust Alignment and Properties}
Dust grain alignment by magnetic fields is the physical mechanism underlying starlight polarization. However, the details of alignment efficiency, grain composition, and size distribution across different interstellar environments are not fully understood. One of DragonflyPol's goals is to compare optical polarization fractions in the $r'$-band with far-infrared polarization measurements from other facilities such as Planck or the James Clerk Maxwell Telescope~\citep[where available;][]{Tahanietal2023}.  
Exploring this ratio enables us to probe how dust alignment efficiency varies between the diffuse and dense regimes traced by optical extinction and thermal emission respectively \citep{HensleyDraine2023}. These comparisons provide constraints on grain alignment mechanisms and dust properties across diverse physical conditions, from dense molecular cloud cores to diffuse high-latitude cirrus. In the longer term, multi-band optical polarimetry with DragonflyPol-like configurations can enable direct study of the wavelength dependence of polarization efficiency described by the Serkowski relation \citep{Serkowskietal1975}. Observing polarization at multiple bands (such as $g'$ and $r'$) and exploring Serkowski relation enable us to put further constraints on grain size and composition, and will further complement submillimeter polarimetry to build a more complete multi-wavelength picture of dust grain properties \citep{HensleyDraine2023}. 

\paragraph{Cosmic Microwave Background Foregrounds}
Accurate characterization of Galactic dust polarization is essential for separating primordial signals from Galactic foreground contamination in cosmic microwave background (CMB) polarization studies, particularly for detecting B-mode polarization from primordial gravitational waves \citep[e.g.,][]{PlanckCollaborationetal2020Cosmology}. 
Wide-field optical polarimetry provides tomographic information on the plane-of-sky magnetic field component along the line of sight, which, when combined with stellar distances from \textit{Gaia}, enables mapping of the three-dimensional distribution of polarizing dust layers contributing to the foreground signal. This three-dimensional view of the foreground dust is a key missing ingredient in current CMB foreground models, which largely treat the polarizing medium as a single layer. DragonflyPol's combination of wide field of view and surface brightness sensitivity makes it well suited to map these foreground structures in diffuse high-latitude regions where CMB experiments are most sensitive.

\paragraph{Three-Dimensional Structure of High-Latitude Diffuse Clouds}
Different studies have explored the 3D spatial structure of some interstellar clouds \citep{Grossschedletal2018, Zuckeretal2021}. However, the 3D structure of diffuse, high-latitude clouds remains poorly constrained. Polarimetry of scattered light may provide an avenue for recovering 3D cloud geometry, as the polarization state of scattered light encodes information about the spatial distribution of the scattering medium itself \citep{Kishimoto1999, Konstantinouetal2022}. This approach is distinct from magnetic field tomography: rather than tracing field geometry along the line of sight, it uses the polarization signature of scattered light to reconstruct where the scattering material is distributed in three dimensions. DragonflyPol's wide field of view and sensitivity to low surface brightness emission make it suited to applying and further developing these techniques across extended high-latitude structures, where the large angular scales involved make narrow-field instruments inefficient. These observations may also be combined with complementary techniques such as Faraday tomography \citep{BrentjensDeBruyn2005, VanEcketal2017} to build a more complete picture of both the spatial structure and physical conditions of diffuse clouds. 

\paragraph{Galactic Magnetic Fields and Large-Scale Structure}
On scales larger than individual clouds, the large-scale structure of the Galactic magnetic field and its relationship to the individual clouds remain incompletely mapped, particularly in the optical. 
Additionally, the magnetic field morphology of the Milky Way's halo has puzzled different communities for decades \citep{Dickeyetal2022}. Stellar polarimetry with DragonflyPol, with its $\sim$5\,deg$^2$ field of view, enables efficient mapping of extended Galactic structures that are inaccessible to narrow-field polarimeters, including loops, shells, superbubbles, and high-latitude filaments. These observations will reveal how large-scale Galactic magnetic fields thread through and shape the diffuse interstellar medium, and how field geometry at large scales connects to the smaller-scale fields threading molecular and diffuse clouds. Combined with existing radio and far-infrared datasets, DragonflyPol observations will contribute to building a coherent multi-wavelength picture of Galactic magnetic field structure across a wide range of physical scales.

\paragraph{Optical Polarimetry of Nearby Galaxies}
The polarization of light from nearby galaxies encodes information about both the magnetic field structure and dust grain properties within those systems. Optical polarimetry traces starlight polarization by aligned dust grains within the galaxy itself, as well as scattered light from dust in the interstellar medium, probing the plane-of-sky magnetic field geometry and dust distribution across the galactic disk and halo. DragonflyPol's $\sim2.8''$ pixel scale and wide field of view make it well suited to mapping polarization across the extended disks and halos of nearby galaxies, where low surface brightness emission dominates. These observations can potentially complement radio polarimetry and far-infrared dust emission maps of nearby systems, contributing to a multi-wavelength picture of magnetic field structure and dust properties in external galaxies \citep{Wisniewskietal2007ApJ}.

\paragraph{Transients and Serendipitous Discoveries}
Polarimetry of transient sources, including supernovae, gamma-ray burst afterglows, and other variable phenomena, can reveal the geometry of the emitting region and the structure of surrounding circumstellar material. While transient science is not a primary focus of DragonflyPol, the instrument's wide field of view makes serendipitous detection of polarized transients possible during routine survey operations. More broadly, wide-field polarimetric surveys of previously unexplored regions of parameter space have historically led to unexpected discoveries, and DragonflyPol's unique combination of field coverage and surface brightness sensitivity may reveal polarized structures with no counterpart in existing total-intensity or narrow-field polarimetric surveys.

\paragraph{Axion-Photon Coupling and Fundamental Physics}
Axion-like particles are predicted to couple to photons, resulting in rotation of the linear polarization angle of linearly polarized light \citep{Fujitaetal2019, Oshimaetal2023}. 
Constraining this effect requires very high-precision polarization angle measurements, as the predicted signal is very subtle. DragonflyPol's image stacking capability (combining data from 48 lenses simultaneously) may offer a pathway toward achieving the polarization angle precision required for such measurements. While current commissioning results establish DragonflyPol as a stable platform for linear polarimetry, achieving the precision necessary for axion-photon coupling searches will require continued refinement of calibration procedures, systematic error characterization, and potentially scaling the array beyond the current 48 lenses, and remains a longer-term goal of the project.

\section{Dragonfly Overview}
\label{sec:overview}
\begin{figure}[t]
\centering
\includegraphics[width=0.48\linewidth]{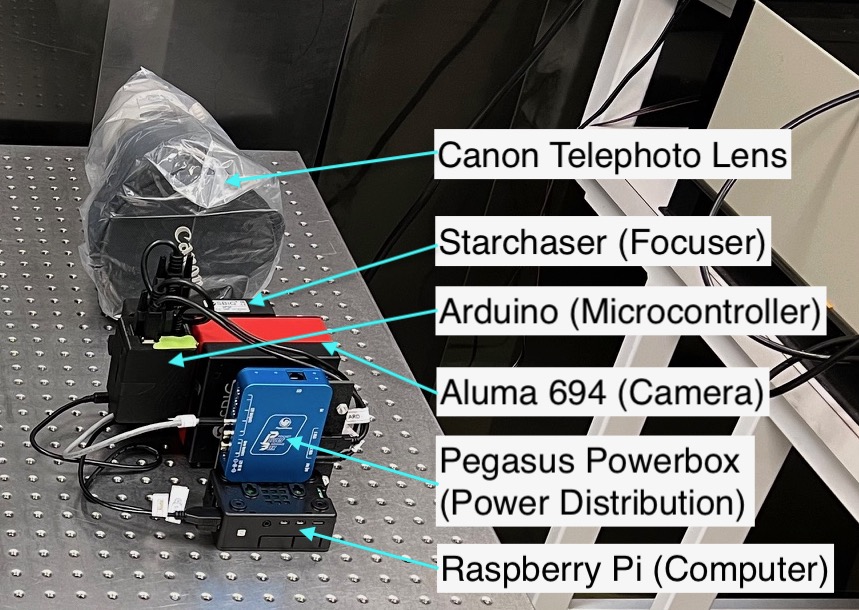}
\includegraphics[width=0.455\linewidth]{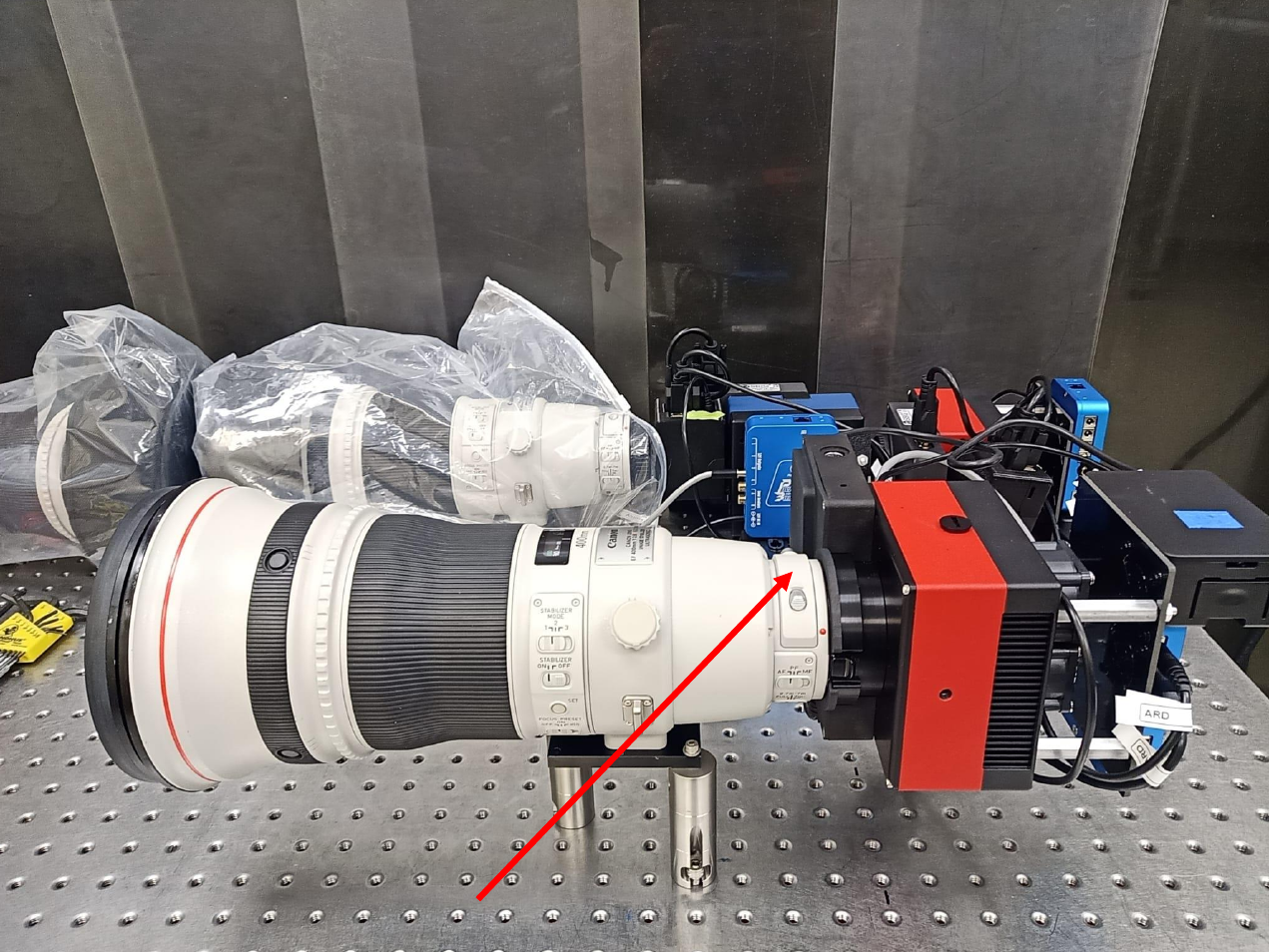}\\
\includegraphics[width=0.41\linewidth]{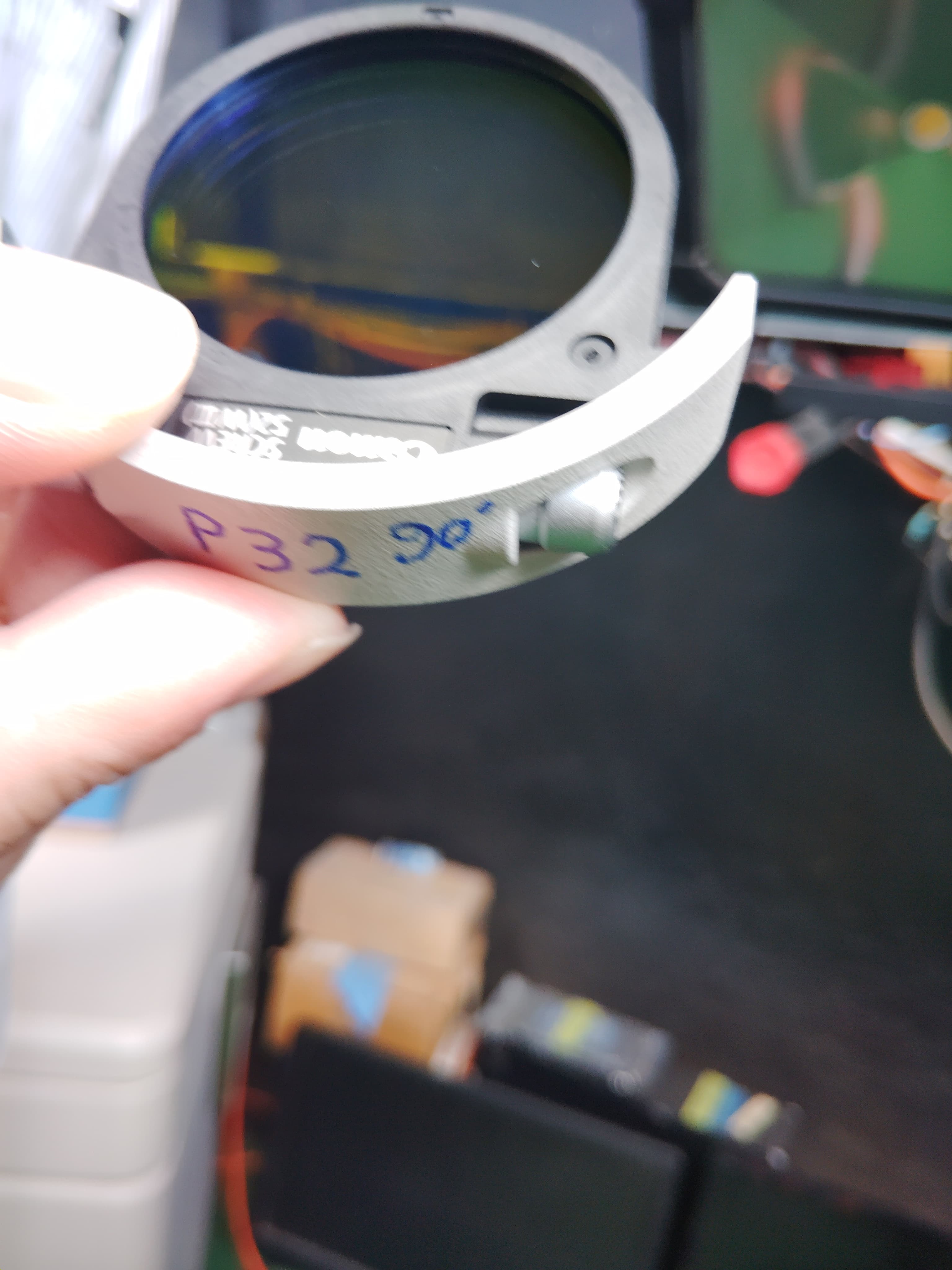}
\includegraphics[width=0.305\linewidth]{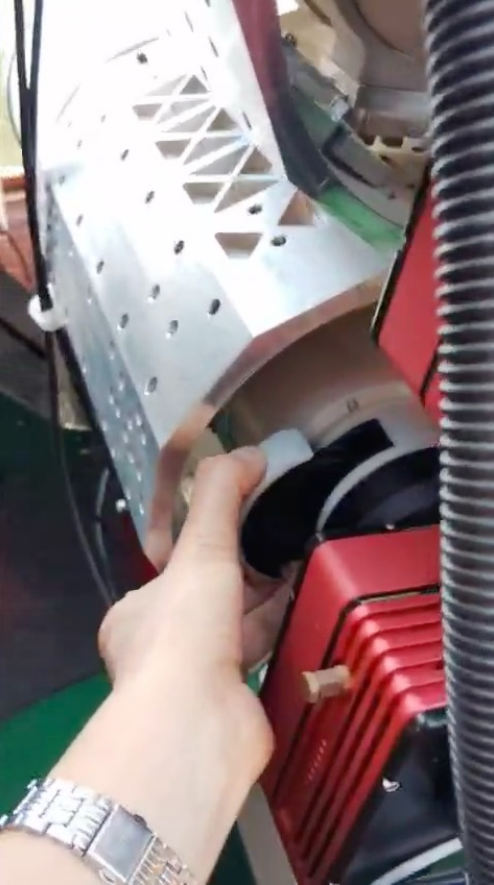}
\caption{DragonflyPol component integration. Top: Canon 400\,mm $f/2.8$ lens, focus electronics, CCD camera, and control microcomputer. Note that the lab-based testing used different electronics (focuser and powerbox) from the on-sky array, but the functionality and command structure is identical. Bottom: Drop-in filter holder containing a bandpass filter and polarizer, and installation at the rear filter slot of the lens.}
\label{fig:lab_components}
\end{figure}

Each of the 48 units comprising the Dragonfly Telephoto Array is a completely independent system with its own optical path, detector, and control electronics. As shown in Figure~\ref{fig:lab_components}, at the heart of each unit is a Canon 400\,mm $f/2.8$ telephoto lens, chosen for its nano-fabricated sub-wavelength anti-reflection coatings that minimize internal light scattering. Focusing is handled by Birger Engineering adapters, which connect the Canon lens to the camera and provide the hardware interface for remote focus commands \citep{AbrahamvanDokkum2014, Danielietal2020}. The primary detectors are SBIG CCD cameras. Power is distributed to all components through DIN-rail mounted DC power supplies. Each unit is controlled by an Intel Compute Stick microcomputer, which sends focus commands to the lens through a Birger focuser module and exposure commands to the detectors through USB.  This modular, self-contained architecture means that each unit can be independently configured, operated, and if necessary reconfigured without affecting the rest of the array. This independent, scalable property is particularly advantageous for polarimetric operations.

Each lens unit accommodates a drop-in filter holder at the rear of the lens, as shown in the top-right panel of Figure~\ref{fig:lab_components}. In this work we designed a system enabling a single filter holder to serve as the mounting structure for both the bandpass filter and polarizer. The Dragonfly team has previously used Baader SDSS/Sloan $g'$ and $r'$ bandpass filters for imaging operations \citep{Danielietal2020}. The laboratory characterization of these filters alongside candidate polarizers is described in Section~\ref{sec:lab}.

\begin{table}[t]
\centering
\caption{Bandpass filters used in laboratory measurements and commissioning \citep{Akitayaetal2024}.}
\begin{tabular}{lccc}
\hline
Filter & Effective wavelength (nm) & Wavelength range (nm) & Manufacturer \\
\hline
SDSS/Sloan $r'$ & 636.4 & 571--703 & Baader \\
SDSS/Sloan $g'$ & 476.4 & 395--559 & Baader \\
\hline
\end{tabular}
\label{tab:filters}
\end{table}

\section{Laboratory Characterization}
\label{sec:lab}

\subsection{Overview, Goals, and Common Laboratory Setup}
During laboratory characterization we primarily focused on (i) selecting a compatible high-performance polarizer and bandpass filter combination for deployment on the array; (ii) measuring contrast ratios and transmission efficiencies of candidate polarizers; (iii) designing a feasible mounting and deployment strategy given overall system constraints; and (iv) determining and marking the transmission axis of each polarizer with repeatable sub-degree precision. To achieve these goals, measurements were conducted in three distinct experimental phases. These phases had different optical components present in the beam path and are described in Sections~\ref{sec:fullsystem}, \ref{sec:PolarizerSelection}, and \ref{sec:scribing}, respectively.

Several components were common to all three phases. The illumination source was a Thorlabs Quartz Tungsten-Halogen lamp (QTH10). The transmission axis orientation of one of the polarizers used in the setup was controlled using a Thorlabs Heavy Duty Rotation Stage (HDR50), driven by a Thorlabs two-channel benchtop stepper motor controller (BSC202). We controlled the motor controller via the Kinesis software interface. A Thorlabs two-axis microscopy joystick console (MJC2) was available for manual polarizer angle adjustment. A Thorlabs iris was placed in the beam path to reduce stray light and improve measurement repeatability. The specific detector, optical path components, and data acquisition method varied by phase and are described in the relevant subsections.

\subsection{Feasibility Tests: Full System Configuration}
\label{sec:fullsystem}
\begin{figure}
\centering
\includegraphics[width=0.2\linewidth]{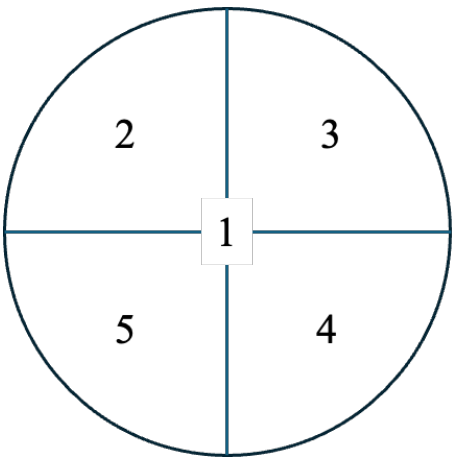}
\caption{The polarizer was divided into four quadrants to examine position-dependent polarization effects. The illumination spot was translated to each quadrant to test whether the polarization axis orientation and contrast ratio varied across the aperture of the polarizer.}
\label{fig:beamTranslation}
\end{figure}

As an initial feasibility test, measurements were carried out with the full optical system (one optical path unit among the 48 units) in place: a Canon 400\,mm $f/2.8$ telephoto lens, a drop-in filter holder containing a candidate polarizer, a Birger Engineering focuser adapter, and an SBIG camera as the detector. In this setup the bandpass filter was located in front of the illumination source along with another polarizer to produce $r'$-band, linearly polarized light. Image acquisition was performed using both MaxIm DL software and FITS image recording through the Raspberry Pi. In this configuration, contrast ratios were measured by comparing median pixel counts in images taken with two polarizers in parallel and orthogonal orientations. The goal in these initial measurements was to examine the feasibility of DragonflyPol, and confining the beam very well was not a priority. These measurements yielded contrast ratios significantly exceeding 100, demonstrating that even with standard commercial polarizers, the system could achieve polarization measurements with better than 1\% precision. Additionally, the following tests were performed:
\begin{enumerate}
    \item The lens was swapped between units to confirm that the lens itself did not alter the polarization axis orientation or contrast ratio. 
    \item Careful pressure was applied on the lens to examine the effect of flexure on polarization. 
    \item The illumination spot was translated from the center to the four quadrants of the lens aperture, as shown in Figure~\ref{fig:beamTranslation}, to test for position-dependent polarization effects. Both raw and noise-subtracted contrast ratios varied by less than 5\% across quadrants, confirming that polarimetric extinction is uniform across the polarizer aperture within measurement uncertainties. 
    \item The polarizer and bandpass filter were inclined at angles of up to $\sim30^\circ$ with respect to the normal of the incident beam. For the Canon 400\,mm $f/2.8$ lens, the half-cone angle of the beam at the filter holder location is $\arctan(1/2f) = \arctan(1/5.6) \approx 10^\circ$, so the tested angles well exceed the maximum angle of incidence expected. At $\sim15^\circ$, already exceeding the half-cone angle, both raw and noise-subtracted contrast ratios changed by less than 9\%, corresponding to a negligible impact on polarimetric sensitivity. At more extreme angles of $\sim30^\circ$, the change increases to less than 20\%. This confirms that polarimetric extinction remains stable within the imaging cone angle of the Canon lenses at the filter holder location.
\end{enumerate}
These results established the feasibility of the DragonflyPol concept and motivated the more systematic characterization described in the following sections.

\subsection{Polarizer Selection}
\label{sec:PolarizerSelection}
\subsubsection{Lens and Photodiode Configuration}
\label{sec:lenspd} 
\begin{figure}[t]
\centering
\includegraphics[width=0.95\linewidth]{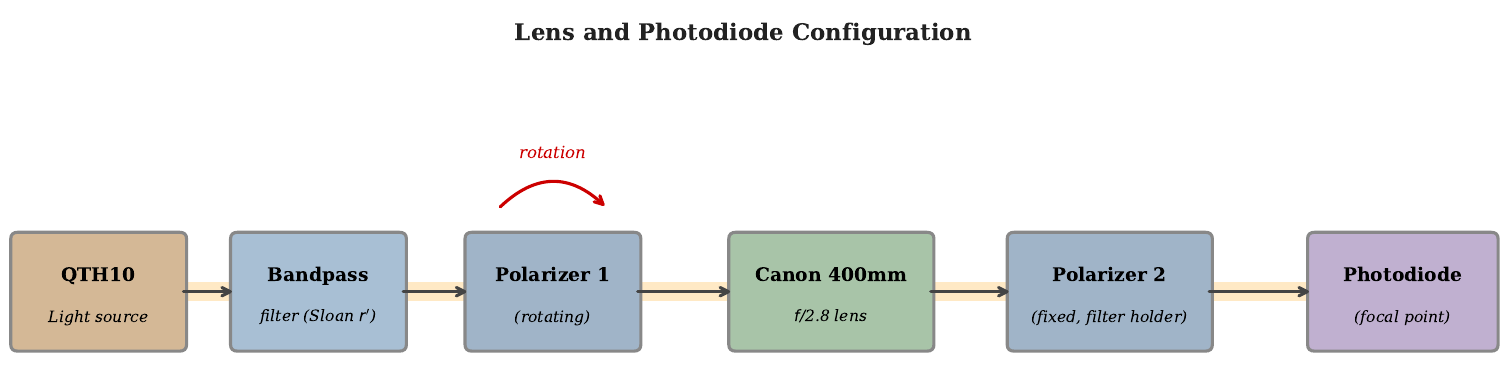}
\caption{Schematic of the laboratory setup used in the lens and photodiode configuration (Section~\ref{sec:lenspd}). The beam from the QTH10 light source passes through a Sloan $r'$ bandpass filter, a rotating polarizer, the Canon 400\,mm $f/2.8$ lens, and a fixed polarizer installed in the drop-in filter holder. The Hamamatsu photodiode is placed at the focal plane of the lens.}
\label{fig:lab_setup_lenspd}
\end{figure}

In the second experimental phase, the camera was removed and replaced by a large-area Si photodiode and transimpedance amplifier (Hamamatsu 8746-01) placed at the focal plane of the Canon lens as shown in Figure~\ref{fig:lab_setup_lenspd}.  This configuration allowed more controlled and quantitative measurements of contrast ratio and transmission efficiency for each candidate polarizer, with the photodiode providing a direct, stable readout of transmitted intensity without the additional complexity of image acquisition and reduction.

Four commercially available polarizer types were evaluated: Canon (circular polarizing filter), Edmund (linear), Hoya (linear), and Tiffen (linear). An additional (significantly cheaper) polarizer sourced online was tested briefly but discarded immediately due to poor optical quality. Although the Canon filter is technically a circular polarizing filter, it consists of a linear polarizer followed by a quarter-wave plate, as one polarizing unit. In our measurements it was used as a linear polarizer by orienting it such that the incident light passes through its linear polarizer component first, before its wave plate. 
Contrast ratios were measured by rotating the second polarizer between parallel and orthogonal orientations relative to the first and recording the photodiode output at each position.  For measuring contrast ratios using two Canon polarizers, 
the two filters were oriented in opposite directions, ensuring the system behaves as two linear polarizers in series. 
Canon and Edmund polarizers were tested in combination with both the Sloan $r'$ and $g'$ bandpass filters, as they both showed higher performance compared to Tiffen and Hoya. Transmission efficiency was measured as the ratio of the photodiode signal with a single polarizer in the beam to the signal with no polarizer present.  

Representative results are as follows. Edmund--Edmund polarizer pairs achieved contrast ratios typically $\gtrsim 2\times10^3$ (noise-subtracted) depending on iris setting and geometry. Canon--Canon and Canon--Edmund combinations achieved contrast ratios of a few$\times10^3$ (noise-subtracted), typically around $3\times10^3$. While Tiffen polarizers exhibited high apparent transmission in some configurations, they showed poor contrast ratios (order $\sim$few) and were not considered further. Hoya polarizers had a higher contrast ratio ($\gtrsim 1000$) compared to Tiffen but lower compared to Edmund and Canon. 

\begin{figure}
\centering
\includegraphics[width=0.286\linewidth]{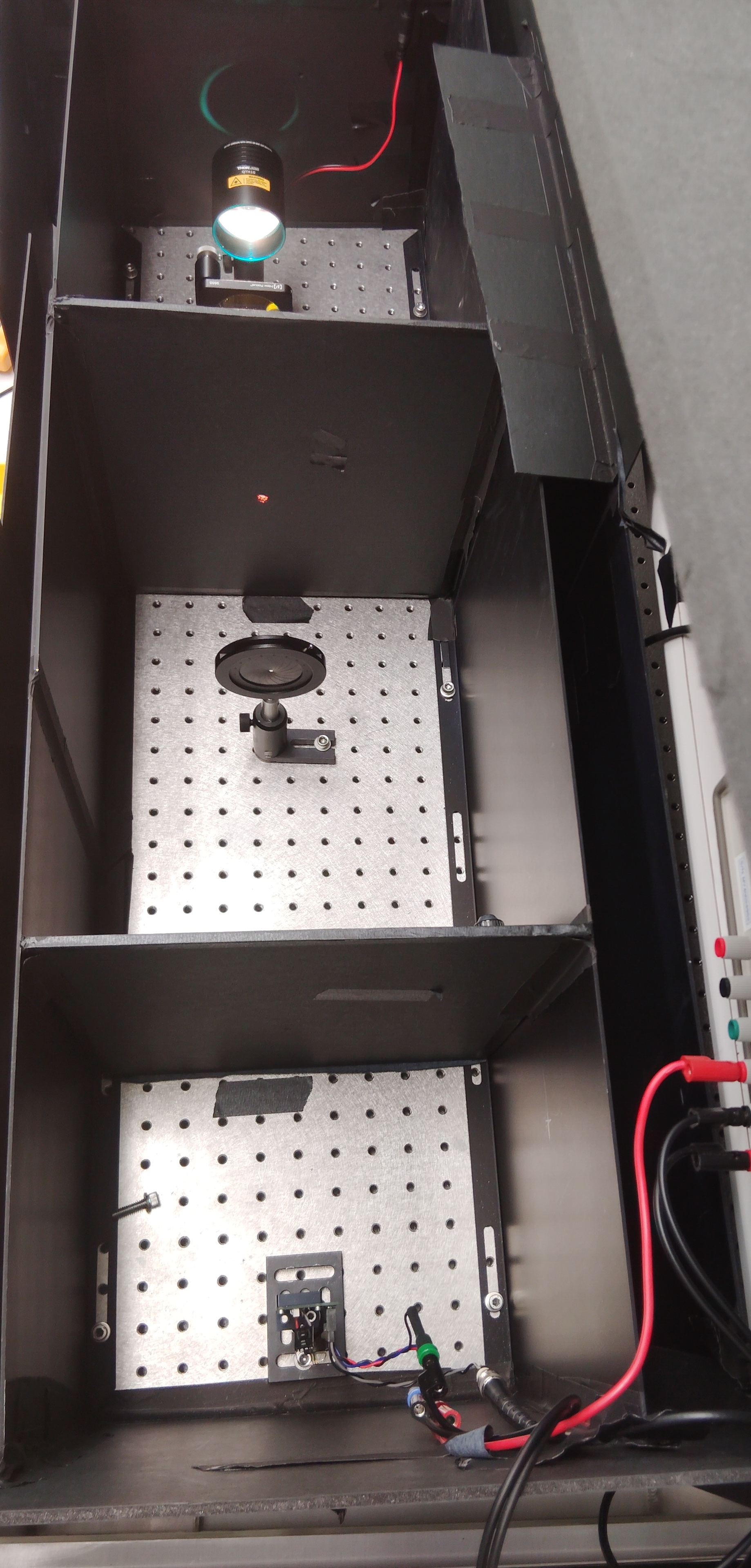} \hspace{0.1cm}
\includegraphics[width=0.45\linewidth]{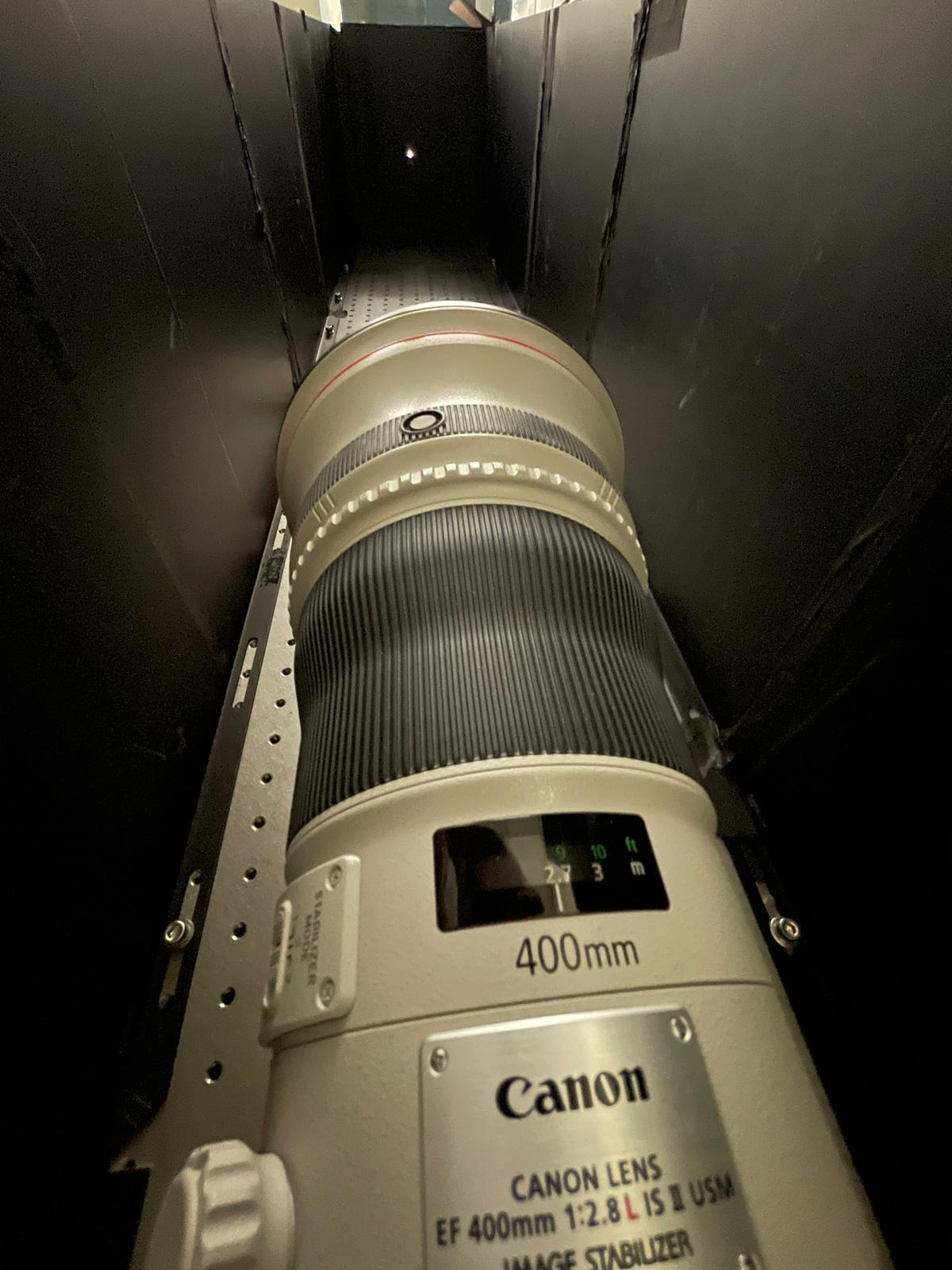}
\caption{Laboratory setup for polarimetric characterization. Left: Setup used for transmission efficiency measurements, showing the illumination source, polarizer, and bandpass filter assembly with photodiode readout. Right: Setup with the Canon 400\,mm $f/2.8$ lens in place. An additional beam-confining wall and iris were typically included in the configuration to reduce stray light, though not shown here.}
\label{fig:transmission}
\end{figure}

Transmission efficiency measurements had a setup shown in the left panel of Figure~\ref{fig:transmission}. Canon polarizers achieved approximately 33\% transmission in the $r'$ band with the Baader SLOAN $r'$ bandpass filter in place. We note that insertion of the bandpass filter reduces overall throughput from 43\% in no-filter configuration to 33\% when we add the bandpass filter, emphasizing the importance of characterizing the polarizers with bandpass filters in place. The reduction in throughput upon insertion of the bandpass filter reflects the broad spectral output of the QTH10 illumination source (400$\sim$2200\,nm), a large fraction of which falls outside the Sloan $r'$ bandpass (571--703\,nm) and is blocked by the filter. 

To test for position-dependent polarization effects, the illumination spot was translated to the four quadrants of the lens aperture (Figure~\ref{fig:beamTranslation}). Contrast ratio and polarization axis orientation were found to be consistent across the aperture within measurement uncertainties. For the photodection amplifier different transimpedance resistance values were used depending on the measurement goal. For contrast ratio measurements, a high resistance (10\,M$\Omega$) was used to maximize sensitivity to the minimum signal in the crossed-polarizer state. For transmission efficiency measurements, a lower resistance (1\,M$\Omega$) was used to prevent saturation. Unless otherwise noted, a single resistance value was maintained throughout a given measurement sequence. Canon circular polarizers were selected for deployment on DragonflyPol based on their combination of high contrast ratio, best transmission efficiency across both bandpass filters, consistent performance across units, and straightforward mechanical integration into the drop-in filter holders of the Canon lenses.

\begin{figure}[t]
\centering
\includegraphics[width=0.7\linewidth]{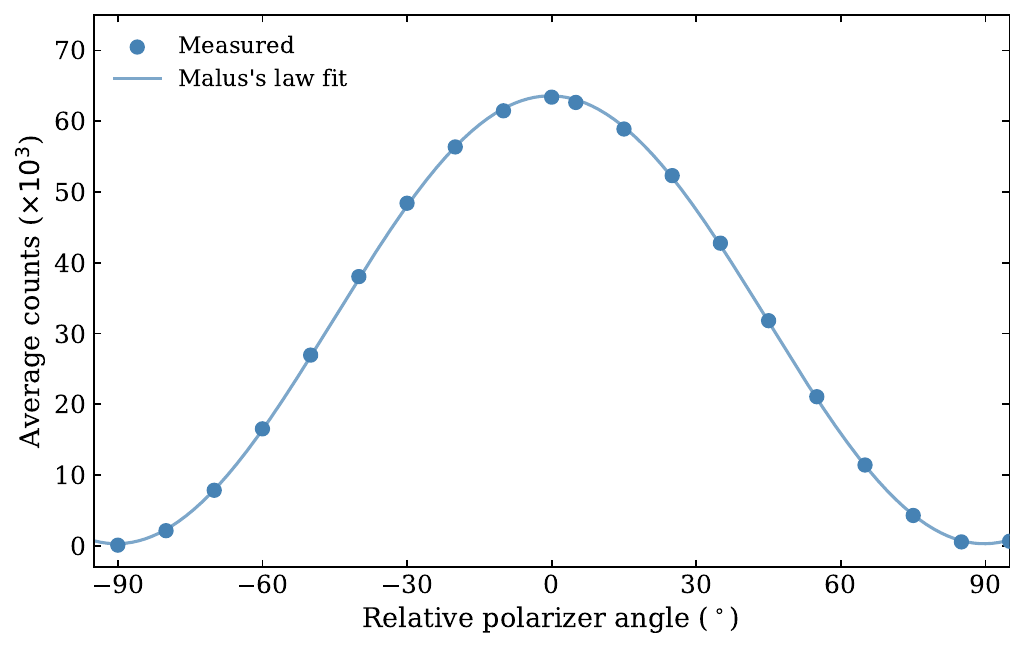}
\caption{Measured photodiode response as a function of the relative angle between two Canon circular polarizers in the lens and photodiode configuration, with the Sloan $r'$ bandpass filter in place. The blue points show the measured average counts within the aperture at each rotation angle, and the solid curve shows the best-fit Malus's law model.}
\label{fig:malus}
\end{figure}

Figure~\ref{fig:malus} shows the measured photodiode response as a function of polarizer rotation angle for a representative Canon polarizer pair, confirming clean Malus's law behavior across the full rotation range. In this setup, one Canon polarizer was placed in front of the lens and rotated through $180^\circ$, and the second was positioned in the drop-in filter holder of the lens.

Figure~\ref{fig:malus} shows the measured photodiode response as a function of polarizer rotation angle for a representative Canon polarizer pair, confirming clean Malus's law behavior across the full rotation range. In this setup, one Canon polarizer was placed in front of the lens and rotated through $180^\circ$, and the second was positioned in the drop-in filter holder of the lens. Although Canon circular polarizing filters each consist of a linear polarizer followed by a quarter-wave plate, the two filters were oriented in opposite directions (the first with the linear polarizer facing the light source and the second with the quarter-wave plate facing the light source) so that light traverses the two quarter-wave plates in opposite order. In this arrangement the system behaves as two linear polarizers in series, following Malus's law as a function of the relative angle between their transmission axes.

\subsubsection{Polarizer and Photodiode Configuration (No Lens)}

\label{sec:nolens}
In the third experimental phase, the Canon lens was removed from the setup to obtain cleaner, more controlled quantitative measurements of polarizer performance without potential contributions from the lens optics. In this configuration the beam path consisted of the QTH10 illumination source, iris, polarizer and bandpass filter assemblies, and the Hamamatsu photodiode directly, with no intervening optics.  Based on the results of the previous phase, which demonstrated that Canon polarizers with the Sloan $r'$ bandpass filter provided the better contrast ratio compared to $g'$ bandpass filter, all subsequent measurements and the final deployed configuration use the $r'$ bandpass filter.

Contrast ratio measurements in this configuration confirmed and refined the results obtained in Section~\ref{sec:lenspd}. Edmund--Edmund polarizer pairs achieved contrast ratios typically $\gtrsim 10^3$ (noise-subtracted) in optimized configurations. Canon--Canon and Canon--Edmund combinations achieved slightly higher contrast ratios. Transmission efficiency measurements with the Sloan $r'$ bandpass filter confirmed Canon polarizers as the best-performing option, achieving approximately 33\% single-polarizer transmission. Without the bandpass filter, transmission efficiencies spanned $\sim$50\%--63\% depending on polarizer type. The high transmission efficiency observed without the bandpass filter reflects the broadband output of the QTH10 illumination source extending beyond the polarizer's optimized wavelength range; at wavelengths outside this range, the polarizer transmits both polarization states more equally, increasing the apparent transmission efficiency without a $r'$ bandpass filter. The observed values may also partly reflect a small intrinsic partial polarization of the illumination source. The illumination source was found to carry a small intrinsic linear polarization at the $\sim$1\%--2\% level. This was identified by comparing measurements with and without a polarizer in the beam and is accounted for in calibration. Within measurement precision, no significant instrumental polarization contributions were identified from the tested bandpass filters or mechanical mounts. 

These measurements confirmed Canon circular polarizers in combination with the Sloan $r'$ bandpass filter as the optimal configuration for DragonflyPol deployment, providing the best combination of contrast ratio, transmission efficiency, and consistency across units.

\subsection{Polarization Axis Marking}
\label{sec:scribing}
\begin{figure}[t]
\centering
\includegraphics[width=0.65\linewidth]{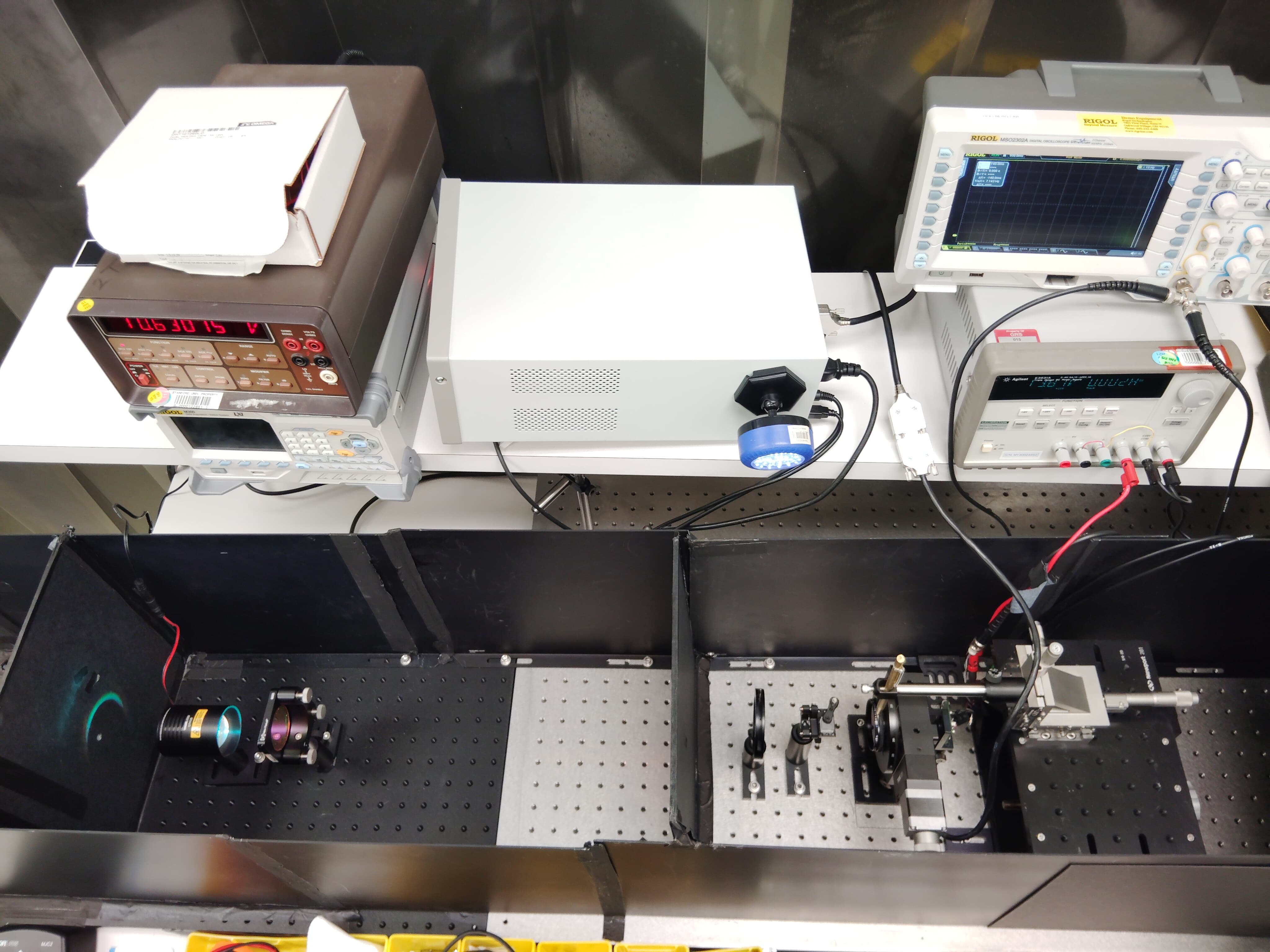}
\caption{Polarization axis scribing station developed for DragonflyPol. The setup enables repeatable transmission axis marking at better than $0.2^\circ$ precision. Components include the QTH10 illumination source, Sloan $r'$ bandpass filter, pinhole, polarizing beamsplitter cube (reference axis), motorized rotation stage, and Hamamatsu photodiode readout.}
\label{fig:scribing_station}
\end{figure}

\begin{figure}[t]
\centering
\includegraphics[width=0.95\linewidth]{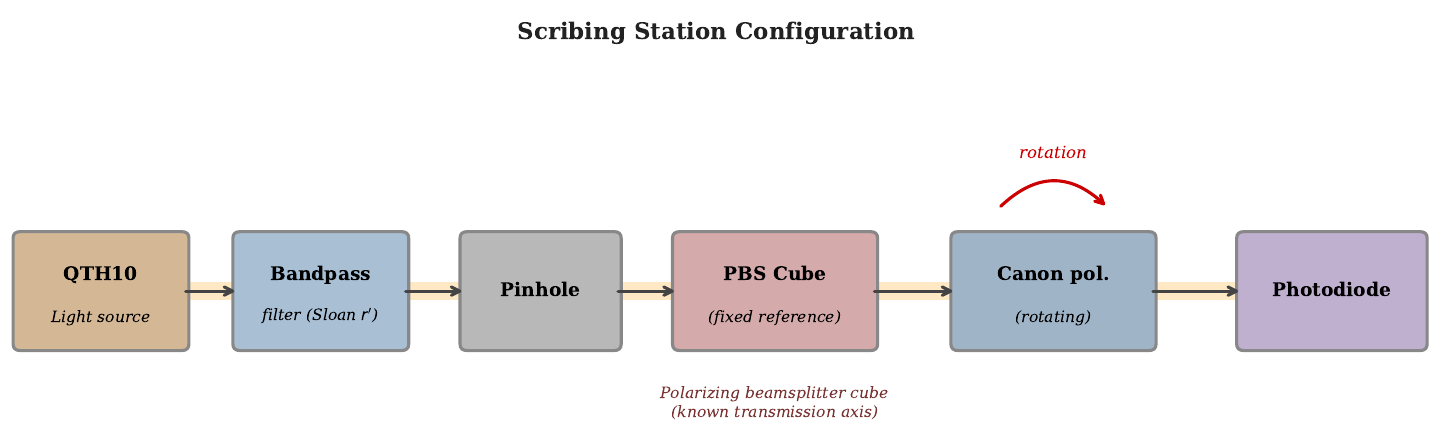}
\caption{Schematic of the polarization axis scribing station (Section~\ref{sec:scribing}). The beam from the QTH10 light source passes through a Sloan $r'$ bandpass filter, a pinhole, and a polarizing beamsplitter cube (PBS) serving as the fixed reference axis. The Canon circular polarizer under characterization is placed after the cube and rotated on the motorized stage; the photodiode records the transmitted intensity as a function of rotation angle, from which the transmission axis is determined.}
\label{fig:lab_setup_scribing}
\end{figure}

Following the selection of Canon circular polarizers for deployment, a dedicated scribing station was developed to determine and mark the transmission axis of each polarizer unit with sub-degree repeatability (Figures~\ref{fig:scribing_station} and \ref{fig:lab_setup_scribing}). Precise knowledge of the transmission axis orientation of each polarizer is essential for accurate Stokes parameter reconstruction.

The scribing station consists of the QTH10 illumination source, a Sloan $r'$ bandpass filter, a pinhole to define the beam, a polarizing beamsplitter cube serving as the reference axis, a motorized rotation stage, and the Hamamatsu photodiode readout. Several reference polarizer configurations were evaluated for establishing the reference axis, including a nanoparticle polarizer with a known transmission axis; the polarizing beamsplitter cube was found to provide the best combination of alignment precision and repeatability and was adopted as the reference standard for all subsequent scribing measurements.

The transmission axis of each polarizer was determined by rotating it on the motorized stage and recording the photodiode output as a function of angle. The resulting intensity curve was fitted to locate the extrema, from which the transmission axis orientation was determined. A dedicated alignment setup was then used to mark the transmission axis orientation on the filter holder and align the polarizer with it.

The Canon polarizer is supplied by the manufacturer with a threaded ring that is carefully screwed into the drop-in filter holder. A machine-shopped protractor was used to mark the filter holder and align and fix the polarizer assembly to the correct position angle during installation into the filter holder. The precision of the polarizer-to-filter-holder alignment is estimated to be better than $1^\circ$, and the scribing repeatability is better than $0.2^\circ$. The Baader Sloan $r'$ bandpass filter is placed between the lens and the polarizer (i.e., light from the sky passes through the lens first, then the bandpass filter, then the polarizer before reaching the camera). The bandpass filter is secured within the filter holder using 3--4 small dots of UV-cured adhesive (JB Weld) applied at the sides, holding it in place without applying mechanical stress to the polarizer. Any residual instrumental polarization introduced by the bonding will be characterized and corrected as part of the calibration process presented in the calibration paper (Tahani et al. in preparation).

\begin{figure}[t]
\centering
\includegraphics[width=0.9\linewidth]{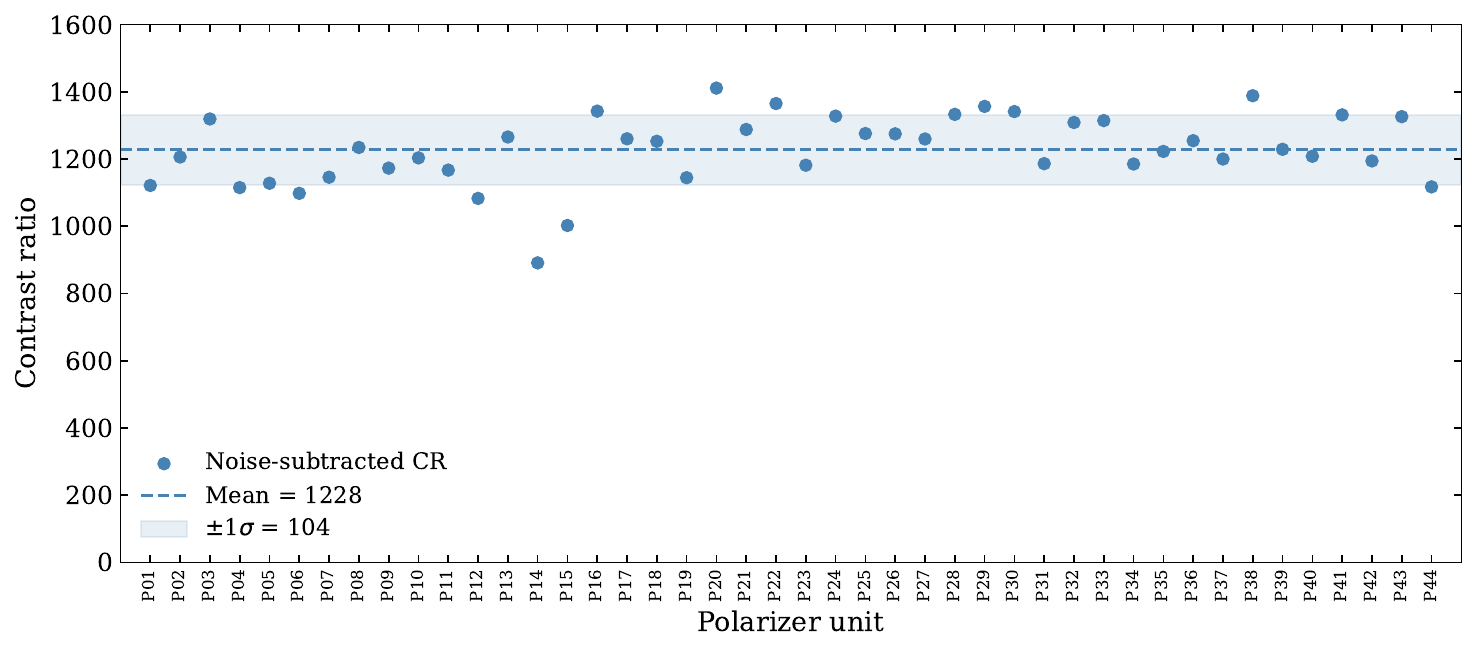}
\caption{Noise-subtracted contrast ratios (blue) for all 44 Canon circular polarizers measured during the scribing characterization session on 2025 August 14--15. The dashed line shows the mean noise-subtracted contrast ratio of $1228$, and the shaded band indicates the $\pm1\sigma$ range ($\sigma = 104$). All units exceed a contrast ratio of 800, well above the minimum threshold established in feasibility tests. Measurements were performed using the Sloan $r'$ bandpass filter and a 10\,M$\Omega$ transimpedance resistance.}
\label{fig:contrast_ratios}
\end{figure}

Table~\ref{tab:scribing} and Figure~\ref{fig:contrast_ratios} present the scribing characterization results for all 44 Canon circular polarizers (P01--P44), measured on 2025 August 14--15. For each polarizer, the transmission axis was first located by finding the intensity minimum on the motorized rotation stage and zeroing the stage at that position. Voltages were then recorded at $0^\circ$, $45^\circ$, $90^\circ$, and $135^\circ$ to verify Malus's law behavior and confirm the axis determination. Here $V_{\rm min}$ is the photodiode voltage at $90^\circ$ (transmission axis minimum), $V_{\rm max}$ is the voltage at $0^\circ$ (transmission axis maximum), and $V_{\rm dark}$ is the dark (box noise) voltage. Raw contrast ratios are computed as $V_{\rm max}/V_{\rm min}$ and noise-subtracted contrast ratios as $(V_{\rm max}-V_{\rm dark})/ (V_{\rm min}-V_{\rm dark})$. The measured $V_{45}/V_{90}$ ratio is $0.498 \pm 0.004$ across all units, consistent with ideal Malus's law behavior to within measurement precision. Raw contrast ratios range from 787 to 1158 with a mean of  $1029 \pm 73$; noise-subtracted contrast ratios range from 891 to 1411 with a mean of $1228 \pm 104$. All units exceed the minimum contrast ratio of 100:1 established in the feasibility tests (Section~\ref{sec:fullsystem}) by more than an order of magnitude. The scribing repeatability is better than $0.2^\circ$ as described above. 

\section{Instrument Configuration, Design, and On-Sky Commissioning}
\label{sec:instrumentCommission}

\subsection{Opto-mechanical Integration and Polarizer Angle Configuration}

\label{sec:polconfig}
\begin{figure}[t]
\centering
\includegraphics[scale=0.5, trim={2cm 2cm 2cm 2cm}, clip]{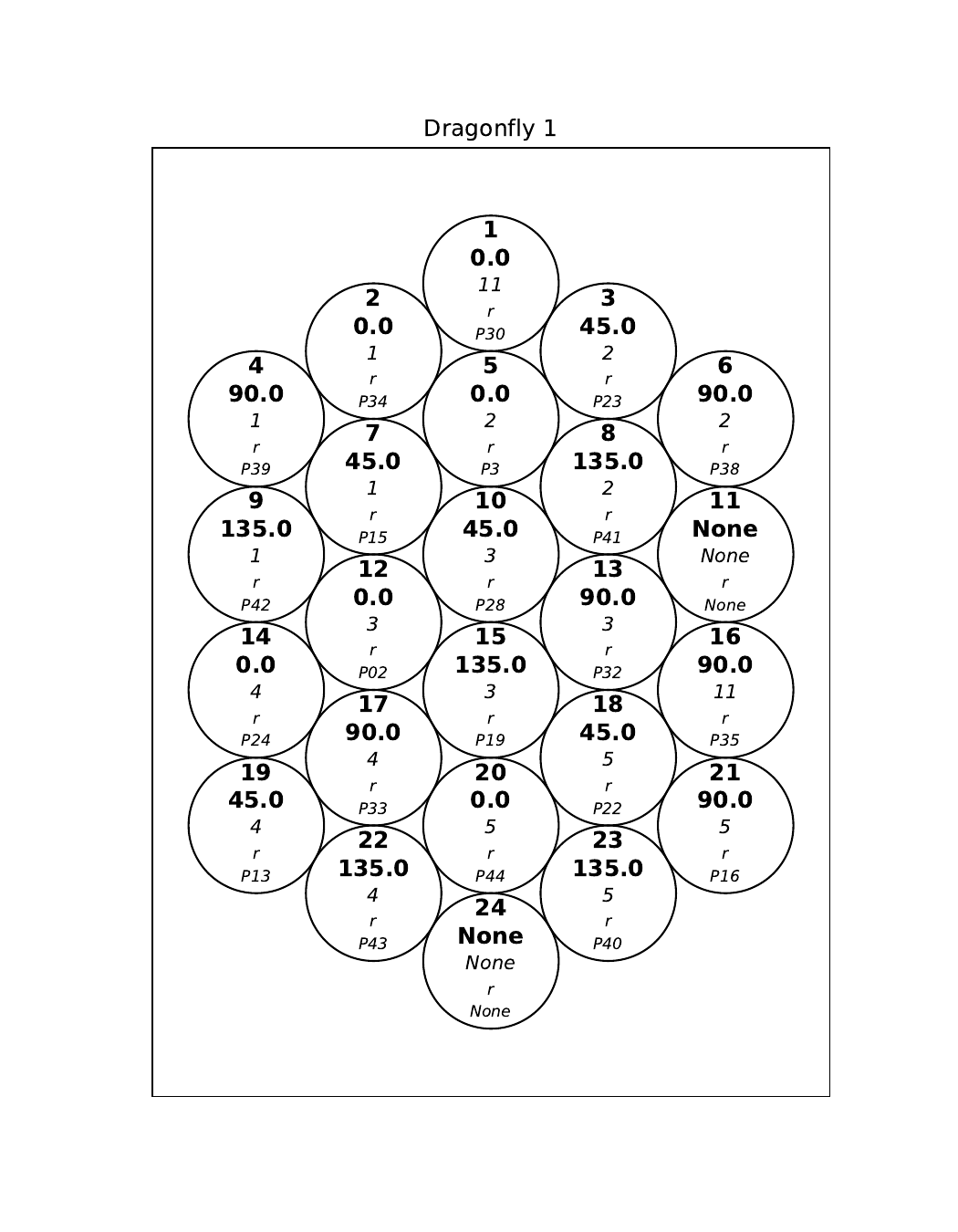}
\includegraphics[scale=0.5, trim={2cm 2cm 2cm 2cm}, clip]{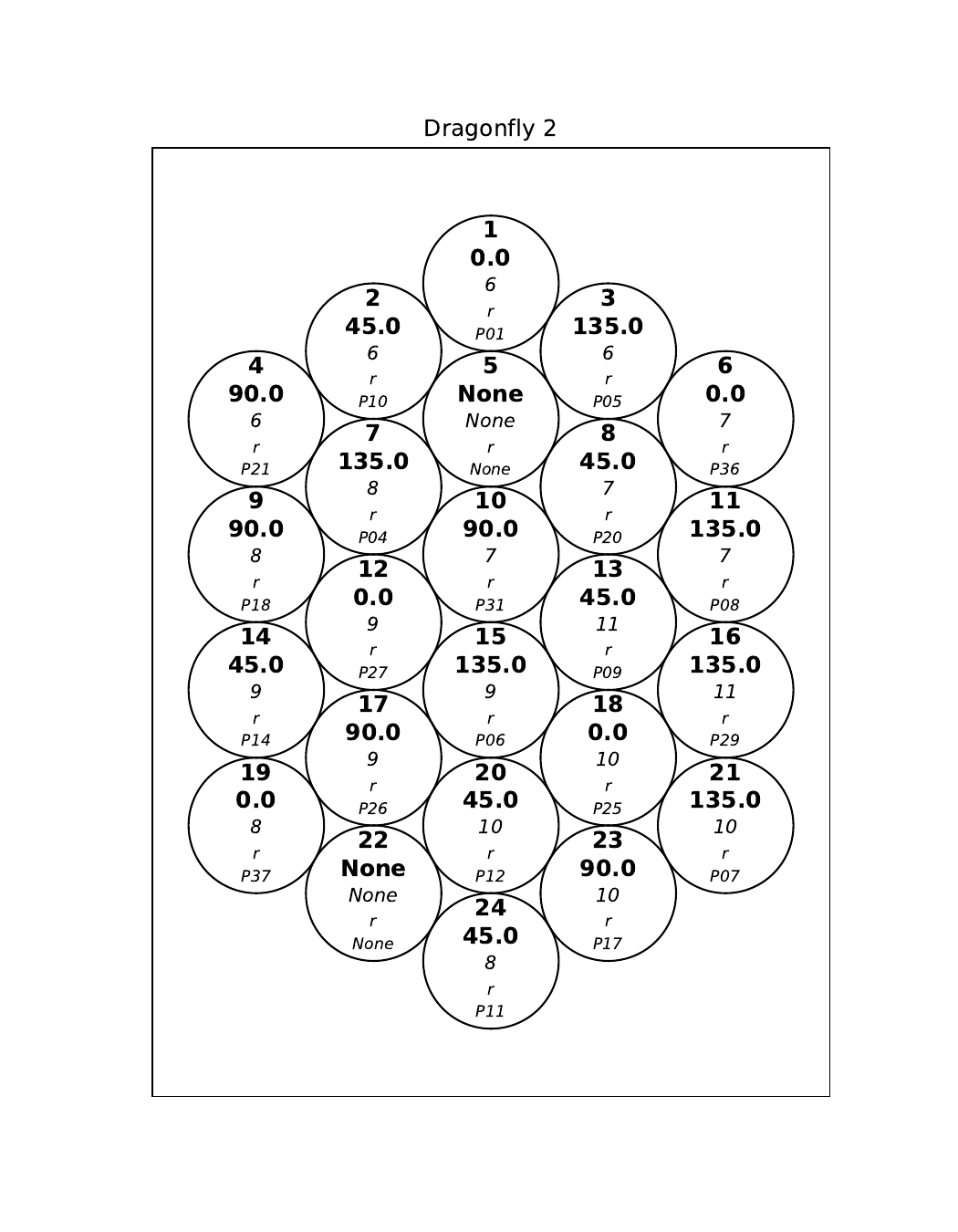}
\caption{Polarizer angle configuration for Mount 1 (left, Dragonfly\,101--124) and Mount 2 (right, Dragonfly\,201--224). Each circle represents one lens--detector unit. For each polarized unit, the bold number indicates the assigned polarization position angle ($0^\circ$, $45^\circ$, $90^\circ$, or $135^\circ$), the italic number indicates the Stokes group assignment (1--11), and the small text indicates the bandpass filter and polarizer assigned number. Units without a polarizer (Dragonfly\,111, 124, 205, 222) are indicated by ``None". Group 11 is the only group spanning both mounts.}
\label{fig:array_configuration}
\end{figure}

The DragonflyPol polarimetric capability is implemented by installing a polarizer and bandpass filter assembly into the drop-in filter holder at the rear of each Canon lens. Each assembly consists of a Canon polarizer and a Baader Sloan $r'$ bandpass filter, fixed together within the filter holder, as described in Section~\ref{sec:scribing}. This approach requires no modification to the Canon lens or camera, and the assemblies can be installed or removed independently on each unit without affecting the rest of the array. Of the 48 lens--detector units, 44 are equipped with polarizer assemblies. The remaining four units (Dragonfly\,111, 124, 205, and 222) carry only an $r'$ bandpass filter without a polarizer, and serve as unpolarized reference channels for monitoring total intensity and tracking sky transparency variations during observations. The 44 polarized units are organized into 11 Stokes groups, each consisting of four lenses assigned fixed polarization position angles of $0^\circ$, $45^\circ$, $90^\circ$, and $135^\circ$ respectively. Within each group, simultaneous imaging at all four polarization angles enables reconstruction of the linear Stokes parameters $Q/I$ and $U/I$, and hence the polarization fraction and position angle, from a single set of exposures. Each polarization angle is represented by exactly 11 lenses across the full array, ensuring a balanced contribution to each Stokes parameter. Groups 1--5 are assigned to Mount 1 (Dragonfly\,101--124) and Groups 6--10 to Mount 2 (Dragonfly\,201--224). Group 11 is distributed across both mounts, with cameras 101 and 116 on Mount 1 and cameras 213 and 216 on Mount 2. This cross-mount assignment was intentional, both to satisfy the constraint that each group must contain exactly one lens at each of the four polarization angles given the physical distribution of available lens positions, and to provide an independent cross-check of the relative calibration between the two mounts. The polarizer angle configuration for both mounts is shown in Figure~\ref{fig:array_configuration}.

\subsection{On-Sky Commissioning}
\label{sec:commissioning}

\begin{figure}[t]
\centering
\begin{tabular}{ccc}
\includegraphics[width=0.3\linewidth]{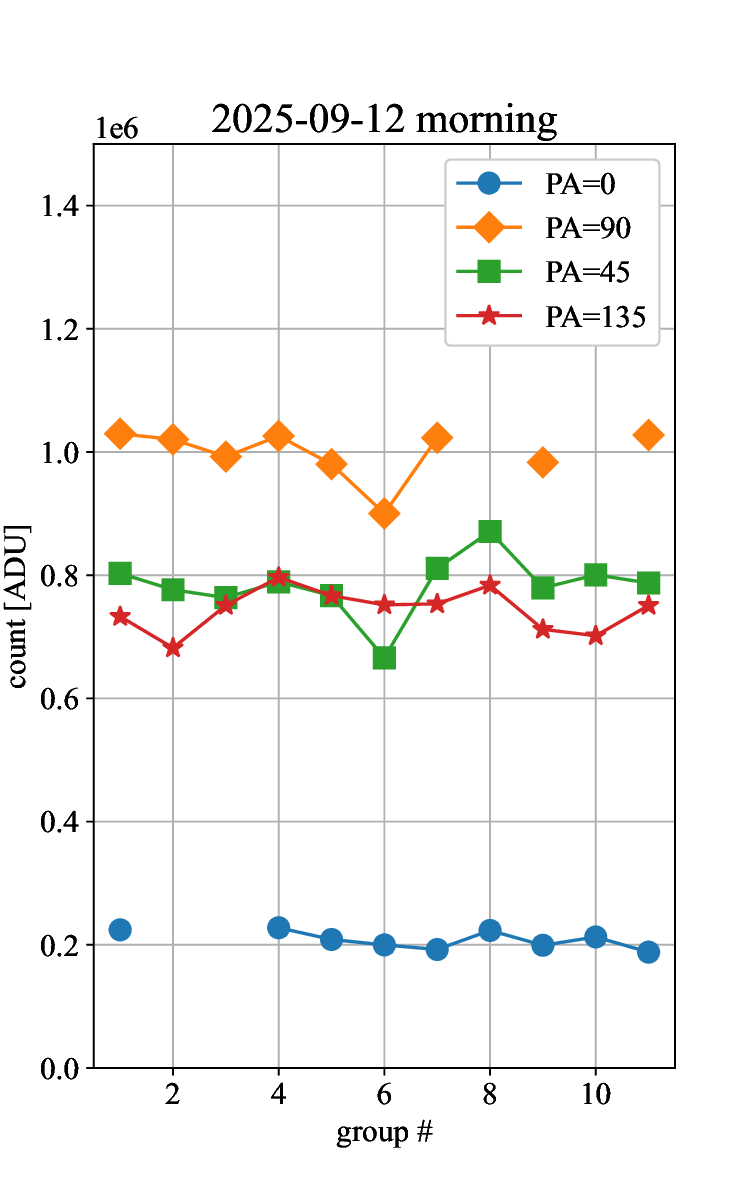} & 
\includegraphics[width=0.3\linewidth]{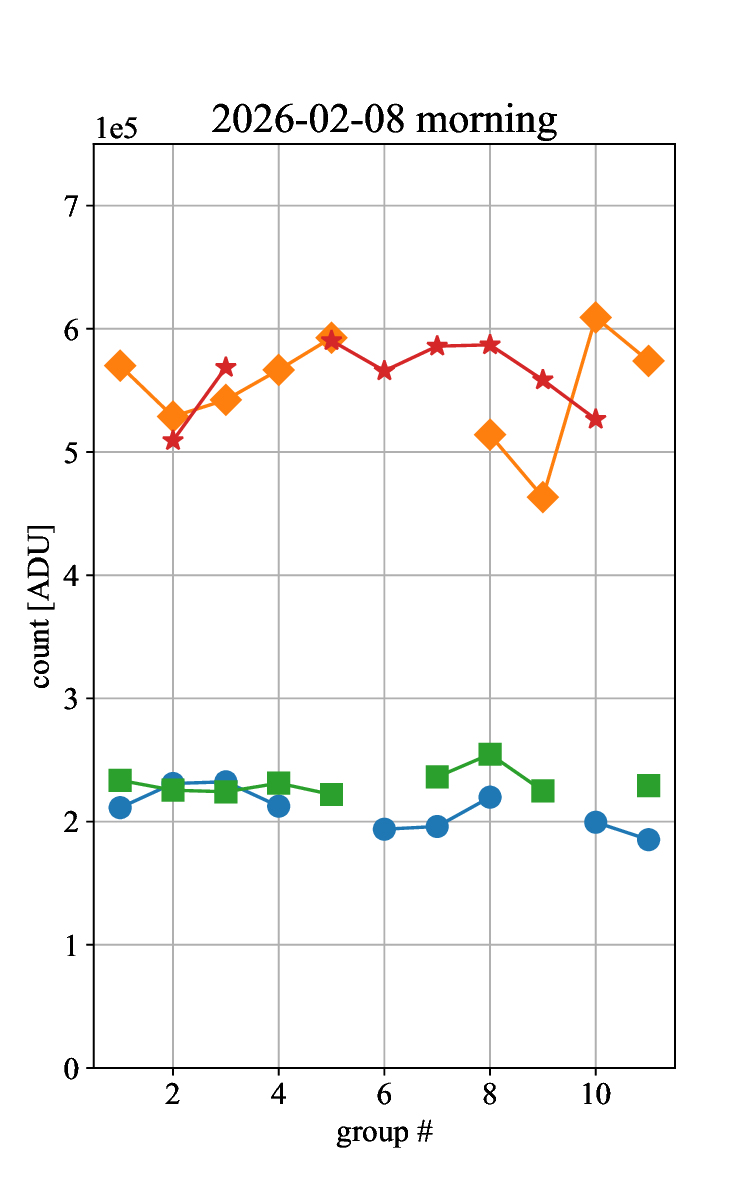} & 
\includegraphics[width=0.3\linewidth]{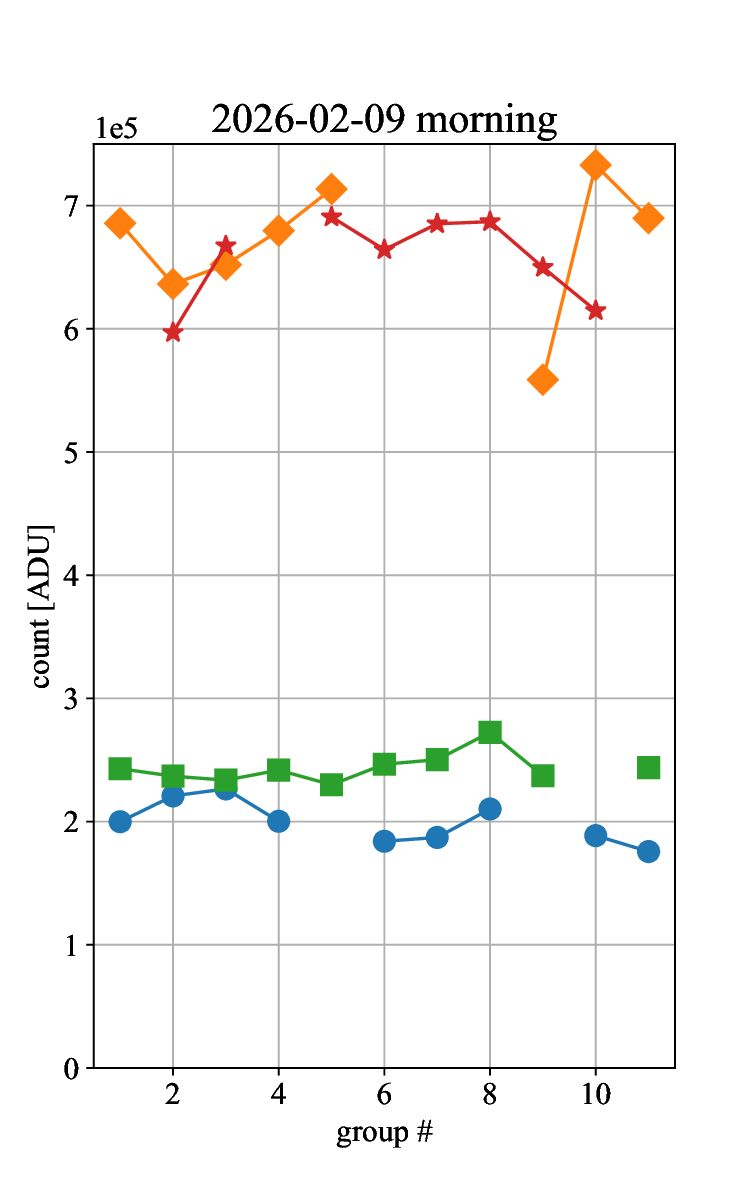} \\
\end{tabular}
\caption{ Total counts in twilight flat fields across the array, grouped by polarization angle ($0^\circ$, $45^\circ$, $90^\circ$, $135^\circ$). Signal levels are broadly stable across units sharing the same polarization orientation. Occasional missing points correspond to dropped frames; a temporary outlier was traced to a momentary malfunction and re-tested subsequently.}
\label{fig:flat_count_plot_rev}
\end{figure}

Commissioning observations were carried out following integration of the polarizer+filter assemblies on the telescope. The primary goals were to validate (i) camera-to-camera throughput stability across the array, (ii) internal consistency across polarization-angle groups, and (iii) successful measurement of a strong, predictable polarization signal from the twilight sky, specifically the linear polarization arising from Rayleigh scattering, whose degree and position angle can be calculated from the observing geometry and compared directly to the measured values.
Twilight flat fields were obtained on three epochs (2025 September 12, 2026 February 8, and 2026 February 9) to evaluate detector and throughput stability across the array. Figure~\ref{fig:flat_count_plot_rev} shows the total counts in twilight flat fields grouped by polarization angle for each epoch. The signal levels are broadly stable across units sharing the same polarization orientation, confirming consistent throughput across the polarized lenses. Occasional missing points correspond to dropped frames or temporary malfunctions that were subsequently re-tested. The systematic difference in signal level between polarization-angle groups is the expected polarimetric response of the strongly polarized twilight sky: lenses whose transmission axes are aligned with the dominant sky polarization direction record the highest counts, while those orthogonal to it record the lowest. The uniformity of signal within each group is the diagnostic of instrumental throughput stability.

The twilight sky provides a strong, predictable linear polarization signal arising from Rayleigh scattering of sunlight in the atmosphere, making it well suited for end-to-end validation of the polarimetric system. Table~\ref{tab:twilight_sky_rev} summarizes the measured Stokes parameters, polarization degree, and position angles for all three epochs, alongside the expected position angles computed from the observing geometry. The measured position angles agree with the expected values to within $\Delta\theta \leq 1.6^\circ$ across all three epochs, confirming accurate polarimetric response of the system. The measured polarization degrees range from 0.62 to 0.72, consistent with prior twilight-sky polarization measurements at similar Sun--target separation angles \citep{Croninetal2005TwilightPol}. A full treatment of the polarimetric calibration performance is presented in the companion calibration paper (Tahani et al. in preparation).

\begin{table}[t]
\centering
\caption{Representative twilight-sky polarization parameters in instrumental coordinates.}
\begin{tabular}{lccc}
\hline\hline
Date (UT) & 2025-09-12 & 2026-02-08 & 2026-02-09 \\
\hline
Stokes $q$(=$Q/I$) (standard deviation)& -0.72 (0.03) & -0.46 (0.05) & -0.55 (0.04)\\
Stokes $u$(=$U/I$) (standard deviation)& +0.03 (0.03) & -0.42 (0.03) & -0.47 (0.04)  \\
Polarization degree & 72\% & 62\% & 72\% \\
Polarization position angle $(\theta_{\mathrm obs})$ & 89.5$^{\circ}$& 111.5$^{\circ}$ & 110.2$^{\circ}$\\
Expected position angle $(\theta_{\mathrm ex})$& 87.9$^{\circ}$ & 111.2$^{\circ}$ &110.8$^{\circ}$\\
$\Delta{\theta}(=\theta_{\mathrm ex}-\theta_{\mathrm obs})$ & -1.6$^{\circ}$ & -0.3$^{\circ}$ & 0.6$^{\circ}$\\
\hline
Separation angle to the Sun & 84.8$^{\circ}$ & 80.8$^{\circ}$ & 80.9$^{\circ}$\\
Altitude of the Sun & -6.7$^{\circ}$ & -6.8$^{\circ}$ & -6.7$^{\circ}$\\
Separation angle to the Moon & 31.1$^{\circ}$ & 54.7$^{\circ}$ &52.3$^{\circ}$ \\
Altitude of the Moon & 31.1$^{\circ}$ & 54.7$^{\circ}$ & 52.3$^{\circ}$ \\
Moon phase & 64.4$^{\circ}$ & 79.2$^{\circ}$ & 90.1$^{\circ}$ \\
\hline
\end{tabular}
\label{tab:twilight_sky_rev}
\end{table}

\section{Discussion and Conclusions}
\label{sec:discussion}
We have presented DragonflyPol, a wide-field optical linear polarimetry capability implemented on the Dragonfly Telephoto Array. DragonflyPol leverages Dragonfly's modular, multi-lens architecture to obtain simultaneous measurements in four linear polarization orientations ($0^\circ$, $45^\circ$, $90^\circ$, and $135^\circ$) across a $\sim5\,deg^2$ field of view. The polarizers are distributed across 44 polarized lens--detector units. The remaining four units serve as unpolarized reference channels.

DragonflyPol commissioning demonstrates that wide-field optical linear polarimetry is achievable with the Dragonfly Telephoto Array using commercial components and a scalable multi-lens architecture. Laboratory characterization confirms that Canon circular polarizers achieve noise-subtracted contrast ratios of $\approx1000$ (mean~=~1228) across all 44 deployed units, with a unit-to-unit standard deviation of 104. These polarizers establish a well-characterized and uniform polarimetric response across the array. The all-refractive optical design minimizes instrumental polarization from internal reflections. These reflections affect mirror-based telescopes and must be carefully modeled and corrected. The scribing procedure yields polarization axis marking repeatability better than $0.2^\circ$, and the protractor-based alignment of polarizers within filter holders achieves a position angle assignment precision of $\sim1^\circ$ across the array. On-sky commissioning confirms throughput stability across polarization groups and successful recovery of the expected twilight-sky Rayleigh scattering signal.

Since first light in September 2025, DragonflyPol has undertaken an active observing program spanning a diverse range of science targets. Observations to date include wide-field polarimetric mapping of the Galactic cirrus clouds, Perseus molecular cloud, Loop~I and Loop~III interstellar structures, diffuse high-latitude fields, and several nearby galaxies including M51, M82, M101, and M27. Standard star observations (including both polarized and unpolarized standards) have been interleaved throughout the program to support ongoing calibration, with a particular focus on standard stars during the first few months of operations. Data reduction, calibration methodology, and science results from this program will be presented in future papers in this series.

DragonflyPol represents a demonstration that wide-field optical polarimetry can be implemented using commercial components, with performance suitable for a broad range of Galactic and extragalactic science goals. The success of this program demonstrates a pathway for implementing novel polarimetric capabilities through modular telescope designs using commercial optical components. This modular approach is significantly more cost-effective than purpose-built polarimetric facilities and readily adaptable to other existing or future modular arrays. The instrument's combination of field coverage, surface brightness sensitivity, and simultaneous multi-angle polarization measurement enables new wide-field optical polarimetric studies of cosmic magnetism, interstellar dust, and the diffuse interstellar medium. 

%% Please use the acknowledgment and contribution environments. This will 
%% be anonomyized when the "anonymous" style option is used. 
\begin{acknowledgments}
We thank the anonymous referee whose careful and constructive comments 
significantly improved this manuscript. We gratefully acknowledge the KIPAC Innovation Grant, which made the initiation of this project possible, and the University of South Carolina startup fund, which supported the purchase of bandpass filters, polarizers, and filter holders for the full 48-unit array. HA was supported by the JSPS KAKENHI Grant-in-Aid for Scientific Research (C) 25K07355. We are grateful for the assistance provided by Alpine Astronomy and Stanford machine shop. We thank David A. Stricker for his invaluable help in procuring the various 
optical and mechanical components used in this project, going above and beyond in his support. We thank the staff at New Mexico Skies Observatory for their outstanding support of this project, particularly Grady Owens. We dedicate a special remembrance to Eugene Malakhov, whose contributions to the success of DragonflyPol at New Mexico Skies were significant and whose presence is deeply missed.
We acknowledge the use of Claude.ai (Anthropic) for manuscript refining assistance and Claude Code for programming support; all outputs were reviewed and verified by the authors.

\end{acknowledgments}

% \begin{contribution}
% %%This section gives authors the space to recognize author contributions. The text inside this environment is NOT counted towards the total word quanta. At a minimum, manuscripts are expected to include this text:

%% But authors are expected to provide more specific details, e.g. 
%%
%%SC was responsible for writing and submitting the manuscript.
%%WWM came up with the initial research concept and edited the manuscript.
%%OTS obtained the funding and edited the manuscript.
%%EBF provided the formal analysis and validation. He also edited the manuscript.
%%GEH Supervised the undergraduates, wrote the software and administers the project github and Zenodo repositories.
%%
%% Authors can use the Contributor Role Taxonomy (CRediT) at
%% https://credit.niso.org
%% for ideas on how write a good statement tailored to their needs.

% \end{contribution}

%% To help institutions obtain information on the effectiveness of their 
%% telescopes the AAS Journals has created a group of keywords for telescope 
%% facilities.
%
%% Following the acknowledgments section, use the following syntax and the
%% \facility{} or \facilities{} macros to list the keywords of facilities used 
%% in the research for the paper.  Each keyword is check against the master 
%% list during copy editing.  Individual instruments can be provided in 
%% parentheses, after the keyword, but they are not verified.
\facilities{New Mexico Skies Observatory.}%HST(STIS), Swift(XRT and UVOT), AAVSO, CTIO:1.3m, CTIO:1.5m, CXO}

%% Similar to \facility{}, there is the optional \software command to allow 
%% authors a place to specify which programs were used during the creation of 
%% the manuscript. Authors should list each code and include either a
%% citation or url to the code inside ()s when available.
% \software{astropy \citep{2013A&A...558A..33A,2018AJ....156..123A,2022ApJ...935..167A},  
%           Cloudy \citep{2013RMxAA..49..137F}, 
%           Source Extractor \citep{1996A&AS..117..393B}
%           }

%% Appendix material should be preceded with a single \appendix command.
%% There should be a \section command for each appendix. Mark appendix
%% subsections with the same markup you use in the main body of the paper.
%%
%% Each Appendix (indicated with \section) will be lettered A, B, C, etc.
%% The equation counter will reset when it encounters the \appendix
%% command and will number appendix equations (A1), (A2), etc. The
%% Figure and Table counter will not reset.

\appendix

\section{Polarizer Scribing Characterization Data}
\label{app:scribing}

In this appendix we present the per-unit scribing characterization data for all 44 Canon circular polarizers (P01--P44) deployed on DragonflyPol, measured on 2025 August 14--15 with the scribing station described in Section~\ref{sec:scribing}. Table~\ref{tab:scribing} lists, for each polarizer, the photodiode voltages at the transmission axis minimum ($V_{\rm min}$) and maximum ($V_{\rm max}$), the dark voltage ($V_{\rm dark}$), and the resulting raw and noise-subtracted contrast ratios, together with the mean and standard deviation across all units. The noise-subtracted contrast ratios from Table~\ref{tab:scribing} are plotted in Figure~\ref{fig:contrast_ratios} in the main text.

\begin{longtable}{lccccc}
\caption{Scribing characterization results for all 44 Canon circular polarizers (P01--P44), measured on 2025 August 14--15 using the polarization axis scribing station described in Section~\ref{sec:scribing}. The transmission axis of each polarizer was zeroed at the intensity minimum before scribing. $V_{\rm min}$ is the photodiode voltage at $90^\circ$ (transmission axis minimum), $V_{\rm max}$ is the voltage at $0^\circ$ (transmission axis maximum), and $V_{\rm dark}$ is the dark (box noise) voltage. Contrast ratios are computed both raw ($V_{\rm max}/V_{\rm min}$) and noise-subtracted ($(V_{\rm max}-V_{\rm dark})/(V_{\rm min}-V_{\rm dark})$). All measurements use the Sloan $r'$ bandpass filter and a 10\,M$\Omega$ transimpedance resistance.}
\label{tab:scribing} \\
\hline
Polarizer & $V_{\rm min}$ (mV) & $V_{\rm max}$ (V) & $V_{\rm dark}$ (mV) & CR (raw) & CR (noise-sub) \\
\hline
\endfirsthead
\hline
Polarizer & $V_{\rm min}$ (mV) & $V_{\rm max}$ (V) & $V_{\rm dark}$ (mV) & CR (raw) & CR (noise-sub) \\
\hline
\endhead
\hline
\endfoot
P01 & 9.48 & 9.027 & 1.43 & 952 & 1121 \\
P02 & 8.94 & 9.057 & 1.43 & 1013 & 1206 \\
P03 & 8.30 & 9.063 & 1.43 & 1092 & 1319 \\
P04 & 9.79 & 9.344 & 1.41 & 954 & 1115 \\
P05 & 9.44 & 9.035 & 1.43 & 957 & 1128 \\
P06 & 9.72 & 9.102 & 1.43 & 936 & 1098 \\
P07 & 9.47 & 9.213 & 1.43 & 973 & 1146 \\
P08 & 8.58 & 8.851 & 1.41 & 1032 & 1234 \\
P09 & 9.01 & 8.914 & 1.41 & 989 & 1173 \\
P10 & 9.08 & 9.230 & 1.41 & 1017 & 1203 \\
P11 & 9.33 & 9.243 & 1.41 & 991 & 1167 \\
P12 & 9.83 & 9.118 & 1.41 & 928 & 1083 \\
P13 & 8.44 & 8.922 & 1.39 & 1057 & 1265 \\
P14 & 11.86 & 9.329 & 1.39 & 787 & 891 \\
P15 & 10.60 & 9.221 & 1.40 & 870 & 1002 \\
P16 & 8.01 & 8.874 & 1.40 & 1108 & 1342 \\
P17 & 8.61 & 9.086 & 1.40 & 1055 & 1260 \\
P18 & 8.43 & 8.820 & 1.39 & 1046 & 1253 \\
P19 & 9.36 & 9.121 & 1.39 & 974 & 1144 \\
P20 & 7.74 & 8.960 & 1.39 & 1158 & 1411 \\
P21 & 8.31 & 8.912 & 1.39 & 1072 & 1288 \\
P22 & 7.83 & 8.791 & 1.39 & 1123 & 1365 \\
P23 & 8.98 & 8.968 & 1.39 & 999 & 1181 \\
P24 & 8.11 & 8.922 & 1.39 & 1100 & 1327 \\
P25 & 8.47 & 9.033 & 1.39 & 1066 & 1276 \\
P26 & 8.41 & 8.950 & 1.39 & 1064 & 1275 \\
P27 & 8.54 & 9.007 & 1.39 & 1055 & 1260 \\
P28 & 7.93 & 8.717 & 1.39 & 1099 & 1333 \\
P29 & 7.93 & 8.872 & 1.39 & 1119 & 1356 \\
P30 & 7.92 & 8.758 & 1.39 & 1106 & 1341 \\
P31 & 9.04 & 9.075 & 1.39 & 1004 & 1186 \\
P32 & 8.11 & 8.795 & 1.39 & 1084 & 1309 \\
P33 & 8.00 & 8.688 & 1.39 & 1086 & 1314 \\
P34 & 8.90 & 8.901 & 1.39 & 1000 & 1185 \\
P35 & 8.72 & 8.962 & 1.39 & 1028 & 1222 \\
P36 & 8.57 & 9.008 & 1.39 & 1051 & 1254 \\
P37 & 8.97 & 9.096 & 1.39 & 1014 & 1200 \\
P38 & 7.75 & 8.830 & 1.39 & 1139 & 1388 \\
P39 & 8.55 & 8.799 & 1.39 & 1029 & 1229 \\
P40 & 8.67 & 8.796 & 1.39 & 1015 & 1208 \\
P41 & 8.00 & 8.800 & 1.39 & 1100 & 1331 \\
P42 & 8.69 & 8.719 & 1.39 & 1003 & 1194 \\
P43 & 8.18 & 8.991 & 1.40 & 1099 & 1326 \\
P44 & 9.61 & 8.971 & 1.58 & 934 & 1117 \\
\hline
\multicolumn{6}{l}{Mean $\pm$ std:} \\
& $8.78\pm0.80$ & $8.975\pm0.162$ & -- & $1029\pm73$ & $1228\pm104$ \\
\end{longtable}

%% For this sample we use BibTeX plus aasjournalv7.bst to generate the
%% the bibliography. The sample7.bib file was populated from ADS. To
%% get the citations to show in the compiled file do the following:
%%
%% pdflatex sample7.tex
%% bibtext sample7
%% pdflatex sample7.tex
%% pdflatex sample7.tex

\bibliography{AllBiblio} {}
\bibliographystyle{aasjournalv7}

%% This command is needed to show the entire author+affiliation list when
%% the collaboration and author truncation commands are used.  It has to
%% go at the end of the manuscript.
%\allauthors

%% Include this line if you are using the \added, \replaced, \deleted
%% commands to see a summary list of all changes at the end of the article.
%\listofchanges

\end{document}